\documentclass[11pt]{article}
\usepackage{amsmath,amsthm,latexsym,amssymb,amsfonts,epsfig}

\makeatletter
\@addtoreset{equation}{section}
\makeatother

\newtheorem{Theorem}{Theorem}[section]
\newtheorem{Definition}{Definition}[section]
\newtheorem{Lemma}{Lemma}[section]
\newtheorem{Corollary}{Corollary}[section]

\def\be{\begin{equation}}
\def\ee{\end{equation}}
\def\ba{\begin{eqnarray}}
\def\ea{\end{eqnarray}}

\def\Nl{{\mathchoice
{\setbox0=\hbox{$\displaystyle\rm N$}\hbox{\hbox to0pt
{\kern0.4\wd0\vrule height0.9\ht0\hss}\box0}}
{\setbox0=\hbox{$\textstyle\rm N$}\hbox{\hbox to0pt
{\kern0.4\wd0\vrule height0.9\ht0\hss}\box0}}
{\setbox0=\hbox{$\scriptstyle\rm N$}\hbox{\hbox to0pt
{\kern0.4\wd0\vrule height0.9\ht0\hss}\box0}}
{\setbox0=\hbox{$\scriptscriptstyle\rm N$}\hbox{\hbox to0pt
{\kern0.4\wd0\vrule height0.9\ht0\hss}\box0}}}}
\def\Zl{{\mathchoice
{\setbox0=\hbox{$\displaystyle\rm Z$}\hbox{\hbox to0pt
{\kern0.4\wd0\vrule height0.9\ht0\hss}\box0}}
{\setbox0=\hbox{$\textstyle\rm Z$}\hbox{\hbox to0pt
{\kern0.4\wd0\vrule height0.9\ht0\hss}\box0}}
{\setbox0=\hbox{$\scriptstyle\rm Z$}\hbox{\hbox to0pt
{\kern0.4\wd0\vrule height0.9\ht0\hss}\box0}}
{\setbox0=\hbox{$\scriptscriptstyle\rm Z$}\hbox{\hbox to0pt
{\kern0.4\wd0\vrule height0.9\ht0\hss}\box0}}}}
\def\Ql{{\mathchoice
{\setbox0=\hbox{$\displaystyle\rm Q$}\hbox{\hbox to0pt
{\kern0.4\wd0\vrule height0.9\ht0\hss}\box0}}
{\setbox0=\hbox{$\textstyle\rm Q$}\hbox{\hbox to0pt
{\kern0.4\wd0\vrule height0.9\ht0\hss}\box0}}
{\setbox0=\hbox{$\scriptstyle\rm Q$}\hbox{\hbox to0pt
{\kern0.4\wd0\vrule height0.9\ht0\hss}\box0}}
{\setbox0=\hbox{$\scriptscriptstyle\rm Q$}\hbox{\hbox to0pt
{\kern0.4\wd0\vrule height0.9\ht0\hss}\box0}}}}
\def\Rl{{\mathchoice
{\setbox0=\hbox{$\displaystyle\rm R$}\hbox{\hbox to0pt
{\kern0.4\wd0\vrule height0.9\ht0\hss}\box0}}
{\setbox0=\hbox{$\textstyle\rm R$}\hbox{\hbox to0pt
{\kern0.4\wd0\vrule height0.9\ht0\hss}\box0}}
{\setbox0=\hbox{$\scriptstyle\rm R$}\hbox{\hbox to0pt
{\kern0.4\wd0\vrule height0.9\ht0\hss}\box0}}
{\setbox0=\hbox{$\scriptscriptstyle\rm R$}\hbox{\hbox to0pt
{\kern0.4\wd0\vrule height0.9\ht0\hss}\box0}}}}
\def\Cl{{\mathchoice
{\setbox0=\hbox{$\displaystyle\rm C$}\hbox{\hbox to0pt
{\kern0.4\wd0\vrule height0.9\ht0\hss}\box0}}
{\setbox0=\hbox{$\textstyle\rm C$}\hbox{\hbox to0pt
{\kern0.4\wd0\vrule height0.9\ht0\hss}\box0}}
{\setbox0=\hbox{$\scriptstyle\rm C$}\hbox{\hbox to0pt
{\kern0.4\wd0\vrule height0.9\ht0\hss}\box0}}
{\setbox0=\hbox{$\scriptscriptstyle\rm C$}\hbox{\hbox to0pt
{\kern0.4\wd0\vrule height0.9\ht0\hss}\box0}}}}
\def\Hl{{\mathchoice
{\setbox0=\hbox{$\displaystyle\rm H$}\hbox{\hbox to0pt
{\kern0.4\wd0\vrule height0.9\ht0\hss}\box0}}
{\setbox0=\hbox{$\textstyle\rm H$}\hbox{\hbox to0pt
{\kern0.4\wd0\vrule height0.9\ht0\hss}\box0}}
{\setbox0=\hbox{$\scriptstyle\rm H$}\hbox{\hbox to0pt
{\kern0.4\wd0\vrule height0.9\ht0\hss}\box0}}
{\setbox0=\hbox{$\scriptscriptstyle\rm H$}\hbox{\hbox to0pt
{\kern0.4\wd0\vrule height0.9\ht0\hss}\box0}}}}
\def\Ol{{\mathchoice
{\setbox0=\hbox{$\displaystyle\rm O$}\hbox{\hbox to0pt
{\kern0.4\wd0\vrule height0.9\ht0\hss}\box0}}
{\setbox0=\hbox{$\textstyle\rm O$}\hbox{\hbox to0pt
{\kern0.4\wd0\vrule height0.9\ht0\hss}\box0}}
{\setbox0=\hbox{$\scriptstyle\rm O$}\hbox{\hbox to0pt
{\kern0.4\wd0\vrule height0.9\ht0\hss}\box0}}
{\setbox0=\hbox{$\scriptscriptstyle\rm O$}\hbox{\hbox to0pt
{\kern0.4\wd0\vrule height0.9\ht0\hss}\box0}}}}
\DeclareMathOperator{\MC}{\boldsymbol{\mathsf{M}}}
\DeclareMathOperator{\MCO}{\boldsymbol{\widehat{\mathsf{M}}}}

\DeclareMathOperator{\MCA}{\boldsymbol{\mathfrak{M}}}
\DeclareMathOperator{\MCW}{\rm Master\;\;Constraint}
\DeclareMathOperator{\MEW}{\rm Master\;\;Equation}
\DeclareMathOperator{\QMCEW}{\rm Quantum\;\;Master\;\;
Constraint\;\;Equation}
\DeclareMathOperator{\MRW}{\rm Master\;\;Relation}
\DeclareMathOperator{\MCOW}{\rm Master\;\;Constraint\;\;
Operator}
\DeclareMathOperator{\MCAW}{\rm Master\;\;Constraint\;\;Algebra}
\DeclareMathOperator{\MCPW}{\rm Master\;\;Constraint\;\;Programme}

\title{Testing the\\ $\MCPW$\\ for Loop Quantum Gravity\\ 
I. General Framework}
\author{B. 
Dittrich\thanks{dittrich@aei.mpg.de, bdittrich@perimeterinstitute.ca}, 
T. Thiemann\thanks{thiemann@aei.mpg.de, tthiemann@perimeterinstitute.ca}\\
\\
Albert Einstein Institut, MPI f. Gravitationsphysik\\
Am M\"uhlenberg 1, 14476 Potsdam, Germany\\
\\
and\\
\\
Perimeter Institute for Theoretical Physics \\
31 Caroline Street North, Waterloo, ON N2L 2Y5, Canada}
\date{{\small Preprint AEI-2004-116}}

\begin{document}

\maketitle

\begin{abstract}
Recently the $\MCPW$ for Loop Quantum Gravity (LQG) was proposed as a 
classically equivalent way to impose the infinite number of Wheeler -- 
DeWitt constraint equations in terms of a single Master 
Equation. While the proposal has some promising abstract features, it was 
until now barely tested in known models.

In this series of five papers we fill this gap, thereby adding confidence 
to the proposal.
We consider a wide range of models with increasingly more complicated 
constraint algebras, beginning with a finite dimensional, Abelean  
algebra of constraint operators which are linear in the momenta and 
ending with an infinite dimensional, non-Abelean algebra of constraint 
operators which closes with structure functions only and which are not 
even polynomial in the momenta. 

In all these models we apply the $\MCPW$ successfully, however, the full 
flexibility of the method must be exploited in order to complete our task.
This shows that the $\MCPW$ has a wide range of applicability but that
there are many, physically interesting subtleties that must be taken care 
of in doing so. In particular, as we will see, that we can 
possibly construct a
$\MCOW$ for a non -- linear, that is, {\it interacting} Quantum Field 
Theory underlines the strength of the 
background independent formulation of LQG. 

In this first paper we prepare the analysis of our test models by 
outlining the general framework of the Master Constraint Programme.
The models themselves will be studied in the remaining four papers.
As a side result we develop the Direct Integral Decomposition (DID)
Programme for solving quantum constraints as an alternative to Refined 
Algebraic Quantization (RAQ).
\end{abstract}

\newpage

\tableofcontents

\section{Introduction}
\label{s1}

The satisfactory implementation of the quantum dynamics of Loop
Quantum Gravity (LQG) (see e.g. the recent reviews \cite{1.1} and 
the forthcoming books \cite{7.2,7.3}) remains the major unresolved problem 
before reliable and falsifiable quantum gravity predictions can be made.
While there has been progress in the formulation of the quantum dynamics
\cite{7.1}, there remain problems to be resolved before the proposal can 
be called satisfactory. These problems have to do with the semiclassical 
limit of the theory and, related to this, the correct implementation of 
Dirac's algebra
of initial value constraints, consisting of the spatial diffeomorphism 
constraints $C_a(x)$ and the Hamiltonian constraints $C(x)$ respectively.
This algebra, sometimes called the {\bf Hypersurface Deformation Algebra}
or {\bf Dirac Algebra}
$\mathfrak{D}$, has the following structure
\ba \label{1.1}
\{\vec{C}(\vec{N}),\vec{C}(\vec{N}')\} &=& \kappa 
\vec{C}({\cal L}_{\vec{N}} \vec{N}')
\nonumber\\
\{\vec{C}(\vec{N}),C(N')\} &=& \kappa 
C({\cal L}_{\vec{N}} N')
\nonumber\\
\{C(N),C(N')\} &=& \kappa 
\vec{C}([dN N'-N dN'] q^{-1})
\ea
Here $\kappa$ is the gravitational coupling constant, $q$ the spatial 
metric of the leaves of $\Sigma_t$ of a foliation of the spacetime 
manifold $M\cong \Rl\times \sigma$, $\cal L$ denotes the Lie derivative 
and we have smeared the constraints properly with test functions, that is,
$\vec{C}(\vec{N})=\int_\sigma d^3x N^a C_a,\;C(N)=\int_\sigma d^3x N C$.
Notice that the appearance of the {\bf structure function} $q^{-1}$ 
appearing in the last relation of (\ref{1.1}) displays $\mathfrak{D}$
as an algebra which is not an (infinite dimensional) Lie algebra.
This makes the representation theory of (\ref{1.1}) so much more 
comlicated than that for infinite dimensional Lie (Super-)Algebras that 
there is almost nothing known about its representation theory. This looks 
like bad news from the outset for any canonical theory of quantum gravity,
such as LQG, which must seek to provide an honest representation of 
(\ref{1.1}).

There is another algebra $\mathfrak{A}$ in canonical quantum gravity which 
plays a central role, the algebra of kinematical (i.e. not gauge 
invariant) observables which is to separate the points of the classical 
phase space so that more complicated composite functions can be expressed 
in terms of (limits of) them. In LQG a very natural choice is given by the 
holonomy -- flux algebra generated by the relations 
\ba \label{1.2}
\{A(e),A(e')\} &=&0
\nonumber\\
\{E_f(S),E_{f'}(S')\} &=&0
\nonumber\\
\{E_f(S),A(e)\} &=& \kappa f^j(S\cap e) A(e_1) \tau_j A(e_2)
\ea
where
\be \label{1.3}
A(e)={\cal P}\exp(\int_e A),\;\;E_f(S)=\int_S f^j (\ast E_j)
\ee
Here we have displayed only the generic cases $S\cap S'=\emptyset$, 
$e=e_1\circ e_2;e_1\cap e_2=e\cap S$, $\tau_j$ is a basis for $su(2)$,
$\ast E$ is the background metric independent two form dual to the vector 
density $E^a_j$ and $A$ denotes an $SU(2)$ connection. The group 
Diff$(\sigma)$ of spatial diffeomorphisms of $\sigma$ acts by a group
of automorphisms 
\be \label{1.3a}
\alpha:\;\mbox{Diff}(\sigma)\to \mbox{Aut}(\mathfrak{A});\;
\varphi\mapsto \alpha_\varphi
\ee
where
\be \label{1.4}
\alpha_\varphi(A(e))=A(\varphi(e)),\;
\alpha_\varphi(E_f(S))=E_{f\circ \varphi^{-1}}(\varphi(S))
\ee
In view of the fact that we want to ultimately construct a physical 
Hilbert space of solutions to the spatial diffeomorphism constraint
and the Hamiltonian constraint it is desirable to have at our disposal 
a cyclic and spatially diffeomorphism invariant representation of 
$\mathfrak{A}$. This is best described by a (half) regular, positive 
linear 
functional
$\omega$ on the $C^\ast-$algebra $\mathfrak{A}$ of the 
corresponding Weyl elements satisfying 
$\omega\circ\alpha_\varphi=\omega$ for all 
$\varphi\in\mbox{Diff}(\sigma)$. The corresponding representation is then 
the GNS representation corresponding to $\omega$. The theory of spatially
diffeomorphism invariant representations of $\mathfrak{A}$ has been 
analyzed in detail in \cite{7.4} with the surprising result that there is 
only one such representation, namely the Ashtekar -- Isham -- Lewandowski 
representation \cite{7.5} that has been used exclusively in LQG for 
a decade already, thereby being justified in retrospect. 

Since the positive linear functional $\omega$ is spatially diffeomorphism 
invariant, general theorems from algebraic QFT \cite{1.6} tell us
that we have a unitary representation of Diff$(\sigma)$ on the GNS 
Hilbert space ${\cal H}_\omega$ defined by 
$U(\varphi) \pi_\omega(a)\Omega_\omega=
\pi_\omega(\alpha_\varphi(a))\Omega_\omega$ where 
$\Omega_\omega$ is the cyclic GNS vacuum and $\pi_\omega$ the GNS 
representation. It turns out \cite{8.2} that this representation is 
necessarily such 
that the one parameter unitary groups $t\mapsto \varphi^{\vec{N}}_t$, 
where 
$\varphi^{\vec{N}}_t$ denotes the diffeomorphisms generated by the 
integral curves
of the vector field $\vec{N}$, {\it are not weakly continuous}. By Stone's 
theorem, this means that the self-adjoint operator corresponding to
$\vec{C}(\vec{N})$ does not exist. This fact presents a major obstacle in 
representing the third relation in (\ref{1.1}) which requires the 
infinitesimal generator (even smeared with operator valued structure 
functions) on the right hand side. The first two relations in 
(\ref{1.1}) can be written in the quantum theory
in terms of finite diffeomorphisms via
\be \label{1.5}
U(\varphi)U(\varphi')=U(\varphi\circ \varphi'),\;\;
U(\varphi)\hat{C}(N)U(\varphi)^{-1}
=\hat{C}(\varphi(N))
\ee
but not so the third relation in (\ref{1.1}). Namly, the third relation  
prevents us from exponentiating the Hamitonian constraints themselves 
which do 
not form (together with the spatial diffeomorphism constraints) a Lie 
Algebra due to appearance of the structure functions.

Another major obstacle is that while the spatial diffeomorphisms form a
subalgebra of $\mathfrak{D}$, they do not form an ideal. Now it turns out 
that in the representation ${\cal H}_\omega$ the Hamiltonian constraints
$\hat{C}(N)$ can be defined only by exploiting their dual action 
on the space of solutions to the spatial diffeomorphism constraint ${\cal 
H}_{Diff}$ \cite{8.2}. However, since spatial diffeomorphisms do not form 
an ideal, one cannot define the operators directly on ${\cal H}_{Diff}$ 
itself since this space is not preserved. More precisely:
A regularized Hamiltonian constraint was defined on 
${\cal H}_{Kin}:={\cal H}_\omega$
and the regulator could be removed using an operator topology 
which exploits the structure of ${\cal H}_{Diff}$. A resulting limit 
operator exists by the axiom of choice but there is a huge regularization 
ambiguity. The commutator 
of two Hamiltonian constraint operators is non-vanishing but annihilates 
diffeomorphism invariant states which is precisely what the third relation 
in (\ref{1.1}) should translate into in the quantum theory if there is no 
anomaly. Although these
methods have been tested successfully in several models (see e.g. the 
fifth reference in \cite{7.1} or \cite{1.8}), the status of the 
Hamiltonian constraint is not entirely satisfactory. For instance, the 
right hand side of the commutator does not obviously look like the 
quantization of the right hand side of the third equation in (\ref{1.1})
so that one can doubt the correctness of the semiclassical limit of the
theory. One could argue that this is because even in the 
classical theory it is a non trivial calculation which 
transforms the Poisson bracket of two Hamiltonian constraints 
into the right hand side of the third line of (\ref{1.1}) and 
that in a semiclassical calculation this can be recovered 
because there one can essentially replace operators and 
commutators by classical functions and Poisson brackets. 
However, nobody has shown that so far. Furthermore, the regularization 
ambiguities are bothersome 
although they will disappear on the physical Hilbert space which 
is the joint kernel of all the constraints. Therefore the overall 
situation is far from being satisfactory.

One could summarize this by saying that the representation theory of 
the kinematical algebra $\mathfrak{A}$ and the hypersurface deformation 
algebra $\mathfrak{D}$ are incompatible and what has been done in 
\cite{7.1} is the best what can be achieved in the present setup. In order 
to make progress, the logical way out is to replace either $\mathfrak{A}$
or $\mathfrak{D}$ by a classically equivalent algebra such that their 
representation theories do become compatible. 

The $\MCPW$ \cite{7.0} 
is a proposal for precisely doing that, it replaces the complicated 
algebra $\mathfrak{D}$ by the much simpler $\MCAW$
$\MCA$. Namely, instead of the 
infinitely many Hamiltonian constraints $C(x)=0,\;x\in \sigma$ we define 
the single $\MCW$
\be \label{1.6}
\MC=\frac{1}{2} \int_\sigma\;d^3x\; \frac{C(x)^2}{\sqrt{\det(q)}(x)}
\ee
where $q_{ab}$ denotes the spatial metric constructed from $E^a_j$
via $\det(q)q^{ab}= E^a_j E^b_k \delta^{jk}$.
The vanishing of all the infinitely many Hamiltonian constraints is 
obviously equivalent to the single $\MEW$ $\MC=0$. Now squaring a 
constraint 
as displayed in (\ref{1.6}) looks like a rather drastic step to do in view 
of the following fact: A weak Dirac Observable $O$ is determined by the 
infinite number of relations $\{C(x),O\}_{\MC=0}=0,\;x\in \sigma$. 
However, the condition $\{\MC,O\}_{\MC=0}=0$ is obviously trivially
satisfied for any $O$ so that the $\MCW$ seems to fail detecting weak 
Dirac 
observables. However, this is not the case, it is easy to see that the 
single $\MRW$
\be \label{1.7}
\{\{\MC,O\},O\}_{\MC=0}=0
\ee
is completely equivalent to the infinite number of relations 
$\{C(x),O\}_{\MC=0}=0,\;x\in \sigma$. Therefore the $\MCW$ encodes 
sufficient information in order to perform the constraint analysis.

The point is now the following simplified constraint algebra, called 
the\\ $\MCAW$ $\MCA$
\ba \label{1.8}
\{\vec{C}(\vec{N}),\vec{C}(\vec{N})\} &=& \kappa\vec{C}({\cal L}_{\vec{N}} 
\vec{N}')
\nonumber\\
\{\vec{C}(\vec{N}),\MC\} &=& 0
\nonumber\\
\{\MC,\MC\} &=& 0
\ea
In other words, since we have carefully divided by $\sqrt{\det(q)}$
in (\ref{1.6}), the integrand is a density of weight one and hence the 
integral is spatially diffeomorphism invariant. Thus now spatial 
diffeomorphisms do form an ideal in the $\MCAW$ and whence the $\MCOW$
must preserve ${\cal H}_{Diff}$. This is precisely what we wanted in order 
to remove the regularization ambiguities mentioned above. Furthermore,
the difficult third relation in (\ref{1.1}) is replaced by the simple 
third relation in (\ref{1.8}) which is a tremendous simplification because 
structure functions are avoided. Hence we do have a chance to make 
progress with the representation theory of the $\MCAW$ $\MCA$.

Of course, squaring a constraint in QFT is dangerous also from the 
perspective of the worsened ultraviolet behaviour of the corresponding 
operator and hence the $\MCPW$ has to be performed with due care.
Moreover, the factor ordering problem will be now much more complex 
and different orderings may very well drastically change the size of the 
physical Hilbert space. It is here where anomalies in the usual framework 
will manifest themselves, hence nothing is ``swept under the rug''.
However, as we will see in the next section, the $\MCPW$ has a chance to 
at 
least complete the canonical quantization programme to the very end, 
{\bf with no further mathematical obstructions on the way}. Whether the 
resulting theory is satisfactory then depends solely on the question 
whether the final physical Hilbert space contains a sufficient number of 
semiclassical states in order to have the original classical theory as its 
classical limit.  

Hence the 
$\MCPW$ is so far only a proposal and is far from granted to be 
successful.
It is the purpose of this series of papers to demonstrate by means 
of a selected
list of models that the $\MCPW$ is flexible enough in order to deal 
successfully with a large number of subtleties, in particular, anomalies,
ultraviolet divergences etc. It also offers an alternative  
to the group averaging programme, also called Refined 
Algebraic Quantization (RAQ), \cite{1.10} from which it differs in two 
important aspects: First of all, RAQ needs
as an additional input the selection of a dense and invariant domain for 
all the constraint operators,  
equipped with a finer topology than that of the Hilbert space into which 
it is 
embedded. On the other hand the $\MCPW$ only uses standard spectral theory 
for normal 
operators on a Hilbert space in order to arrive at a direct integral 
decomposition (DID) of the Hilbert space. The physical Hilbert space 
is then the induced zero eigenvalue ``subspace''. That subspace is however
only known up to structures of measure zero and in order to fix the 
remaining ambiguities, additional physical input is needed, namely 
that the induced Hilbert space carries a self adjoint representation of 
the Dirac observables. We will show however that, even without using 
further physical input, the amount of ambiguity for DID is smaller
than for RAQ. The second difference is that 
RAQ, at least so far, cannot rigorously deal with constraint algebras 
which involve 
non-trivial structure functions since the group averaging really requires 
an honest (Lie) group structure. In contrast, the $\MCPW$ does not 
draw 
an essential distinction between the case with structure constants and 
structure functions respectively. Finally the $\MCPW$ is very flexible 
in the sense that for a given set of constraints there is an infinite 
number of associated $\MCW$ functionals which are classically all 
equivalent but which have different quantizations. One can exploit that 
freedom in order to avoid, e.g., ultraviolet problems and factor ordering 
problems as we will see.\\
\\
The present paper is organized as follows:\\
\\

In section two we briefly review the $\MCPW$ from \cite{7.0} for a general 
theory.

Section three develops the general theory of the direct integral 
decomposition (DID) method
for solving quantum constraints. Most of this is standard spectral 
theory for possibly unbounded self -- adjoint operators and will be 
familiar to experts. More precisely, we recall the spectral theorem 
for unbounded self -- adjoint operators in its projection valued measure 
and functional calculus form, give the abstract definition of a 
direct integral, display the direct integral resolution of the spectral 
projections of self -- adjoint operators (functional model), recall how 
to split a Hilbert space into a direct sum such that the respective 
restrictions of the operator has pure point or continuous spectrum (which 
will be important for our applications), connect direct integral 
representations with constraint quantization, derive the explicit action 
of strong Dirac observables on the physical Hilbert space and finally 
compare RAQ and DID methods.    

In section four we describe, for readers not interested in these 
mathematical details displayed in section three, a brief algorithm for 
how to arrive at the 
physical Hilbert space given a self adjoint constraint operator by the 
Direct Integral Decomposition Technique (DID).
Reading that section will be sufficient for readers who are just 
interested in the application of the formalism. 

We conclude in section five and anticipate some of the results of our 
companion papers \cite{II,III,IV,V}.

\section{Review of the Master Constraint Programme}
\label{s2}

We briefly review the $\MCPW$. For more details the reader is referred to 
\cite{7.0}.

Let be given a phase space $\cal M$ with real valued, first class 
constraint functions 
$C_I(y):\;{\cal M}\to \Rl;\;m\mapsto [C_I(y)](m)$ on $\cal M$. 
Here we let take $I\in{\cal I}$ take discrete values while $y\in Y$
belongs to some continuous index set. To be more specific, $Y$ is supposed 
to be a measurable space and we choose a measure $\nu$ on $Y$. Then we 
consider the fiducial Hilbert space 
$\mathfrak{h}:=L_2(X,d\mu)^{|{\cal I}|}$ with inner 
product 
\be \label{2.1}
<u,v>_{\mathfrak{h}}=\int_Y \; d\nu(x)\; \sum_{I\in {\cal I}}
\overline{u_I(y)} v_I(y)
\ee
Finally we choose a positive -- operator valued function 
${\cal M}\to {\cal L}_+(\mathfrak{h});\;m\mapsto K(m)$ where 
${\cal L}_+(\mathfrak{h})$ denotes the cone of positive linear operators 
on $\mathfrak{h}$.
\begin{Definition} \label{def2.1} ~~\\
The $\MCW$ for the system of constraints $m\mapsto 
[C_I(x)](m)$ corresponding to the choice $\nu$ of a measure on $Y$ and 
the operator valued function $m\mapsto K(m)$ is defined by
\be \label{2.2}
\MC(m)=\frac{1}{2} <C(m),K(m)\cdot C(m)>_{\mathfrak{h}}  
\ee
\end{Definition}
Of course $\nu,K$ must be chosen in such a way that (\ref{2.2}) converges 
and that it defines a differentiable function on $\cal M$, but apart from 
that the definition of a $\MCW$ allows a great deal of flexibility which 
we 
will exploit in the examples to be discussed. It is clear that the 
infinite number of constraint equations $C_I(y)=0$ for a.a. $y\in Y$ and 
all $I\in {\cal I}$ is equivalent with the single $\MEW$ $\MC=0$ so that
classically all admissable choices of $\nu,K$ are equivalent.  

Notice that we have explicitly allowed $\cal M$ to be infinite 
dimensional. In case that we have only a finite dimensional phase space,
simply drop the structures $y,Y,\nu$ from the construction.

We compute for any function $O\in C^2({\cal M})$ that 
\be \label{2.3} 
\{\{O,\MC\},O\}_{\MC=0}=[<\{O,C\},K\cdot \{O,C\}>_{\mathfrak{h}}]_{\MC=0}
\ee
hence the $\MRW$ $\{\{O,\MC\},O\}_{\MC=0}=0$ is equivalent with 
$\{O,C_I(y)\}_{\MC=0}=0$ for a.a. $y\in Y$ and $I\in {\cal I}$.
Among the set of all weak Dirac observables satisfying the $\MRW$ the 
strong
Dirac observables form a subset. These are those twice differentiable 
functions on $\cal M$ satisfying $\{O,\MC\}\equiv 0$ 
identically\footnote{Notice that this does not automatically imply that
$\{O,C_I(y)\}=0$ identically for all $y,I$, however, it implies 
$\{O,C_I(y)\}_{\MCO=0}=0$.} on all of
$\cal M$. They can be found as follows: Let $t\mapsto \alpha^{\MC}_t$
be the one-parameter group of automorphisms of $\cal M$ defined by time 
evolution with respect to $\MC$. Then the {\bf ergodic mean} of
$O\in C^\infty({\cal M})$ 
\be \label{2.4}
[O]:=\lim_{T\to\infty} \frac{1}{2T} \int_{-T}^T\; dt\; \alpha^{\MC}_t(O)
\ee
has a good chance to be a strong Dirac observable if twice differentiable
as one can see by formally commuting the integral with the Poisson bracket 
with respect to $\MC$. In order that the limit in (\ref{2.5}) is 
non-trivial, the integral must actually diverge. Using l' Hospital's 
theorem we therefore find that if (\ref{2.4}) converges and the integral
diverges (the limit being an expression of the form $\infty/\infty$) then 
it equals 
\be \label{2.5}
[O]:=\lim_{T\to\infty} 
\frac{1}{2}[\alpha^{\MC}_T(O)+\alpha^{\MC}_{-T}(O)]
\ee
which is a great simplification because, while one can often compute the 
time evolution $\alpha^{\MC}_t$ for a bounded function $O$ (for bounded 
functions the integral will typically diverge linearly in $T$ so that
the limit exists), doing the integral is impossible in most cases.
Hence we see that the $\MCPW$ even provides some insight into the 
structure 
of the classical Dirac observables for the system under 
consideration.

Now we come to the quantum theory. We assume that a judicious choice of
$\nu,K$ has resulted in a positive, self-adjoint operator $\MCO$ on some 
kinematical Hilbert space ${\cal H}_{Kin}$ which we assume to be 
separable. Following (a slight modification of) a proposal due to Klauder 
\cite{2.1}, if zero is not in the 
spectrum of $\MCO$ then compute the 
finite, positive number $\lambda_0:=\inf(\sigma(\MCO))$ and redefine   
$\MCO$ by $\MCO-\lambda_0\mbox{id}_{{\cal H}_{Kin}}$. Here we assume 
that $\lambda_0$ vanishes in the $\hbar\to 0$ limit so that the modified 
operator still qualifies as a quantization of $\MC$. This is justified in 
all examples encountered so far where $\lambda_0$ is usually related to 
some reordering of the operator. More generally, it might be 
necessary to subtract a ``normal ordering operator'' 
$\hat{\lambda}_0$ (so that the resulting operator is still positive) which 
is 
supposed to vanish in the $\hbar\to 0$ limit. See e.g. \cite{8.3} for an 
example where free quantum fields are coupled to 
gravity which could be looked at as a model with second class 
constraints in analogy to the scond example in \cite{II} and the usual
infinite normal ordering constant becomes a densely defined operator.
Hence in what follows we assume w.l.g. that $0\in \sigma(\MCO)$.

Under these circumstances we can {\bf completely solve the 
$\QMCEW$ $\MCO=0$ and explicitly provide the physical Hilbert space and 
its 
physical inner product}. Namely, as it is well known \cite{2.2}
the Hilbert space ${\cal H}_{Kin}$ is unitarily equivalent to a direct 
integral 
\be \label{2.6}
{\cal H}_{Kin}\cong \int_{\Rl^+}^\oplus\;d\mu(x)\;
{\cal H}^\oplus_{Kin}(x)
\ee
where $\mu$ is a so-called spectral measure and 
${\cal H}^\oplus_{Kin}(x)$ is a separable Hilbert space with 
inner product induced from ${\cal H}_{Kin}$. This simply follows from 
spectral theory. The operator $\MCO$ acts on 
${\cal H}^\oplus_{Kin}(x)$ by multiplication by $x$, hence the 
physical Hilbert space is simply given by
\be \label{2.7}
{\cal H}_{Phys}={\cal H}^\oplus_{Kin}(0)
\ee
Strong Quantum Dirac Observables can be constructed in analogy to 
(\ref{2.4}), (\ref{2.5}), namely for a given bounded operator on 
${\cal H}_{Kin}$ we define, if the uniform limit exists
\be \label{2.8}
\widehat{[O]}:=\lim_{T\to\infty} 
\frac{1}{2}[U(T)\hat{O}U(T)^{-1}+U(T)^{-1}\hat{O}U(T)]
\ee
where
\be \label{2.9}
U(t)=e^{it\MCO}
\ee
is the unitary evolution operator corresponding to the self-adjoint 
$\MCO$ via Stone's theorem. One must check whether the spectral 
projections of the bounded operator (\ref{2.8}) commute with those of 
$\MCO$ but if they do then $\widehat{[O]}$ defines a strong quantum Dirac 
observable. Notice, however, that in the case of interest (structure 
functions) strong Dirac observables are not very interesting and weak
Dirac observables can only be constructed by using (\ref{2.8}) by using 
judicious $\hat{O}$ (it has to be invariant under all constraints but 
one). For a systematic procedure to construct weak Dirac 
classical and quantum observables see \cite{BD}.\\
\\
This concludes our sketch of the general theory. We will 
now describe precisely how to arrive at (\ref{2.6}) and (\ref{2.7}).
In particular, there are certain choices to be made and we will state 
precisely how physical predictions will depend on those choices. The 
result is that the presentation of (\ref{2.6}) is actually unique (up to
unitary equivalence) but (\ref{2.7}) can be fixed only by using additional
physical input. We will outline in detail what that input is in order to 
make (\ref{2.7}) essentially as unique as it can possibly be. Readers
not interested in those details who just want to apply DID can skip the 
next section and jump immediately to section \ref{s9} where we summarize 
our findings.

\section{General Framework for the Master Constraint Programme}
\label{s0}

The Master Constraint Programme makes extensive use of the spectral theory 
for self -- adjoint operators, their invariants up to unitary equivalence 
and their functional models, that is, the realization as multiplication 
operators on a direct integral Hilbert space. This theory is of course 
well known in mathematical physics but we feel that it is worthwhile 
reviewing it here so that one has a compact account of the relevant theory
together with the essential proofs at one's disposal. The proofs are also
instructive because one actually learns how the method works in detail.
Specialists can safely skip this section, except for section \ref{s0.5}
where the direct integral decomposition (DID) theory is connected with 
constraint 
quantization and section \ref{s0.7} where DID is compared with RAQ. 
Practitioners not interested 
in the mathematical details can immediately jump to section four where 
we simply summarize in algorithmic form the contents of this section.\\
\\
This section is subdivided as follows:\\ 

First of all we recall the spectral 
theorem for self -- adjoint operators and how to construct the associated 
projection valued measures (p.v.m.). 

Next we define direct integral Hilbert spaces and 
their associated p.v.m. These could also be called bundles 
of Hilbert spaces with fibres ${\cal H}_x$ whose dimension 
may 
vary as 
$x$ varies over a measurable space 
$(X,{\cal B})$ 
where $X$ is some set (the base) and $\cal B$ a 
$\sigma-$algebra\footnote{A $\sigma-$algebra $\cal B$ on a set $X$ is a 
collection of subsets of $X$ which contains $X$ and the empty set 
$\emptyset$ and is closed under countable unions and intersections. The 
members $B\in {\cal B}$ are called measurable sets.}
over $X$, together with a measure\footnote{A measure $\mu$  
on a measurable space $(X,{\cal B})$ is a countably additive, positive 
set function $\mu:\;{\cal B}\to \Rl^+\cup\{\infty\}$, that is, if 
$B_n,\;n=1,2,..$ are mutually disjoint measurable sets then 
$\mu(\cup_n B_n)=\sum_n \mu(B_n)$.} $\mu$. We require that the 
${\cal H}_x$ are all separable, that $X$ is $\sigma-$finite (is a 
countable union of measurable sets each of which has finite $\mu$ measure)
and that $\cal B$ is separable\footnote{I.e. there is a countable 
collection $\cal C$ of measurable sets in $\cal B$, called a base, so 
that for each 
$B\in {\cal B}$ and each $\epsilon >0$ there is a $B_0\in {\cal C}$ 
such that $\mu([B-B_0]\cup[B_0-B])<\epsilon$.}. It turns out that 
direct integrals of Hilbert spaces over the same $(X,{\cal B})$ 
together with their spectral projections are 
unitarily equivalent if and only 
if 1. the associated measures are equivalent\footnote{
For two measures $\mu,\nu$ on $(X,{\cal B})$ we say that $\nu$ is 
absolutely continuous with respect to $\mu$ if $\mu(B)=0$ implies 
$\nu(B)=0$. Mutually absolutely continuous measures are called 
eqivalent.} and 2. the dimension functions $N(x):=\dim({\cal H}_x)$ 
coincide $\mu-$a.e.\footnote{A property holds on $X$ $\mu-$a.e. (almost 
everywhere) if it is violated at most on measurable sets $B$ of vanishing 
$\mu$ measure, that is, $\mu(B)=0$.} 

Next we connect the first and second part by showing that for each 
self -- adjoint operator $a$ on a Hilbert space ${\cal H}$ we find a 
direct integral representation for its spectral projections. 
The role of $(X,{\cal B})$ is here taken by 
$(\sigma(a)\subset \Rl,{\cal B}_{Borel})$ where $\sigma(a)$ denotes the 
spectrum of $a$\footnote{The spectrum of a densely defined and closable
(the adjoint $a^\dagger$ is also densely defined) operator $a$ on a 
Hilbert space 
$\cal H$ is the set of complex 
numbers $\lambda$ such that $a-\lambda 1_{{\cal H}}$ does not have a 
bounded inverse. For self-adjoint operators (that is, $a$ and $a^\dagger$
have the same domain of definition and $a=a^\dagger$) the spectrum is a 
subset of the real 
line.} and ${\cal B}:={\cal B}_{Borel}$ is the natural Borel 
$\sigma-$algebra on 
$\Rl$\footnote{This is the smallest $\sigma-$algebra containing 
all the open sets of $\Rl$ with respect to its natural metric 
topology.}. Not only will 
we give a constructive procedure for how to do that, but also we will show 
that the choices that one has to make within that procedure lead to 
unitarily equivalent representations. Furthermore, we show that all 
operators $b$ commuting with $a$ (that is, the corresponding spectral 
projections commute) are fibre preserving, that is, they induce operators 
$b(x)$ on all ${\cal H}_x$ which are self -- adjoint on ${\cal H}_x$ if 
and only if $b$ is self -- adjoint.   

Then we have to connect this with constraint quantization. If zero 
belongs to the spectrum of $a$ then we would like to choose 
${\cal H}_{phys}:={\cal H}_{x=0}$ as the physical Hilbert space selected 
by the constraint $a=0$. However, here we have to add physical input  
since the set $\{x\}$ is of $\mu-$measure zero provided that 
$x$ is not in the pure point spectrum of $a$. Hence, if $x=0$ does not lie 
in the point spectrum then we are free to set ${\cal H}_x=\{0\}$ without 
affecting the unitary equivalence with the direct integral representation. 
Moreover, if $x=0$ is an 
eigenvalue embedded into the continuous spectrum then the direct integral 
Hilbert space ${\cal H}_{x=0}$ is granted to correspond to the zero 
eigenvectors only while generalized eigenvectors corresponding 
to the continuous spectrum are easily missed. Both features are of course 
unacceptable and we demonstrate how to repair this by adding an additional
requirement which is motivated by the concrete physical examples studied 
so far where our procedure gives the correct results. In order to do this 
properly we have to connect this with the measure theoretic origin of the 
pure point and continuous spectrum respectively which we briefly recall as 
well.

Next we show explicitly how the direct integral decomposition (DID) 
automatically leads 
to an induced self-adjoint representation on the physical Hilbert space of 
strong self adjoint Dirac observable operators.

Last but not least we establish how DID relates 
to the programme of refined algebraic quantization (RAQ). Notice that RAQ
has actually two implementations: A heuristic version, called group 
averaging, and a rigorous version, using Rigged Hilbert Spaces. We show 
that there is no universally applicable group averaging procedure and that
the theory of Rigged Hilbert Spaces uses more structural input than DID
needs. Moreover, DID can deal with structure functions in contrast to 
RAQ.\\
\\
Our exposition is based 
on relevant parts of \cite{2.2,Birman,RS} which should be consulted for 
further information.

\subsection{Spectral Theorem, Projection Valued Measures, Spectral 
Projections, Functional Calculus}
\label{s0.1}

\begin{Definition} \label{def0.1} ~\\
Let $(X,{\cal B})$ be a measurable space and ${\cal H}$ a Hilbert space.
A function $E$ from $\cal B$ into the set of projection operators on 
${\cal H}$ is called a projection valued measure (p.v.m.) provided it 
satisfies \\
1. $E(\cup_{n=1}^\infty B_n)=\sum_{n=1}^\infty E(B_n)$ for mutually 
disjoint 
$B_n\in {\cal B}$.\\
2. $E(X)=1_{{\cal H}}$
\end{Definition}
From the projection property $E(B)^2=E(B)$ one easily derives 
$E(B_1) E(B_2)=E(B_1\cap B_2)$, $E(\emptyset)=0$ and 
$E(B_1)\le E(B_2)$ for $B_1\subset B_2$ where for two projections 
$P_1\le P_2$ means that $P_1{\cal H}\subset P_2{\cal H}$.

Given a p.v.m. $E$ and a unit vector $\Omega\in {\cal H}$ we 
define the spectral measure 
\be \label{0.1}
\mu_\Omega(B):=<\Omega,E(B)\Omega>_{{\cal H}}=:\int_B d\mu_\Omega(x)
\ee
That this defines indeed a positive, normalized, $\sigma-$additive set 
function is easily verified from definition \ref{def0.1}. Using the 
polarization identity 
\ba \label{0.2}
<\Omega_1,E(B) \Omega_2> &=& \frac{1}{4}[
<(\Omega_1+\Omega_2),E(B)(\Omega_1+\Omega_2)>
-<(\Omega_1-\Omega_2),E(B)(\Omega_1-\Omega_2)>
\\
&& -i<(\Omega_1+i\Omega_2),E(B)(\Omega_1+i\Omega_2)>
+i<(\Omega_1-i\Omega_2),E(B)(\Omega_1-i\Omega_2)>]
\nonumber
\ea
we may define the complex measures 
$\mu_{\Omega_1,\Omega_2}(B)=<\Omega_1,E(B)\Omega_2>$ as well.

Given a 
measurable complex valued function\footnote{A function $f:\;X\to Y$ from a 
measurable 
space $X$ to a topological space $Y$ is said 
to be measurable if the set of points $\{x\in X;\;f(x)\in I\}$ is 
measurable for every open set $I\subset Y$.} $f$ on $X$ we may approximate 
it pointwise by simple functions of the form $f_N(x)=\sum_{n=1}^N z_n 
\chi_{B_n}(x)$ 
where $\chi_B$ denotes the chracteristic function of the set 
$B\in {\cal B}$ and $z_n\in \Cl$ (that is, we find a sequence such that 
$\lim_{N\to \infty} f_N(x)=f(x)$ where the rate of convergence may depend 
on $x$, see e.g. \cite{Rudin}). Associate to $f_N$ the operator 
$f_N(E):=\sum_{n=1}^N z_n 
E(B_n)$. Then 
\be \label{0.3}
<\Omega_1,f_N(E)\Omega_2>=
\sum_{n=1}^N z_n <\Omega_1,E(B_n)\Omega_2>
=\sum_{n=1}^N z_n \int_X \chi_{B_n}(x) d\mu_{\Omega_1,\Omega_2}(x)
=\int_X f_N(x) d\mu_{\Omega_1,\Omega_2}(x)
\ee
Passing to the limit\footnote{The interchange of the limit and the 
integral is justified by the Lebesgue dominated convergence theorem, see 
e.g. \cite{Rudin}.} we find \be \label{0.4}
<\Omega_1,f(E)\Omega_2>=\int_X f(x) d\mu_{\Omega_1,\Omega_2}(x)
\ee
for every measurable function $f$. By using the the notation
\be \label{0.5}
d\mu_{\Omega_1,\Omega_2}(x)
=:d<\Omega_1,E(x)\Omega_2>
=:<\Omega_1,dE(x)\Omega_2>
\ee
one writes (\ref{0.4}) often in the form 
\be \label{0.6}
f(E)=\int_X f(x) dE(x)
\ee
whose precise meaning is given by (\ref{0.4}). 

Let now $a$ be a, not necessarily bounded, self -- adjoint operator on 
${\cal H}$. This means that 1. $a$ is densely defined on a domain $D(a)$,
2. that it is symmetric, i.e. $D(a)\subset D(a^\dagger)$ and $a^\dagger=a$
on $D(a)$ and that 3. actually $D(a^\dagger)=D(a)$. Here 
$D(a^\dagger)=\{\psi\in {\cal H};\;\sup_{0\not=\psi'\in D(a)}
|<\psi,a \psi'>|/||\psi'||<\infty\}$. Then the famous spectral theorem 
holds.
\begin{Theorem} \label{th0.1} ~\\
Let $a$ be a self -- adjoint operator on a Hilbert space. Then there 
exists a p.v.m. $E$ on the measurable space 
$(\Rl,{\cal B}_{Borel})$ such that 
\be \label{0.7}
a=\int_{\Rl}\; x\; dE(x)
\ee
where the domain of integration can be restricted to the 
spectrum $\sigma(a)$.
\end{Theorem}
In order to construct $E$ from $a$ we notice that for each measurable, 
bounded set $B$ the function $\chi_B$ has support in a closed, finite 
interval containing $B$, therefore it can be approximated pointwise by 
polynomials due to the Weierstrass theorem. On the other hand, the 
function $x$ can be approximated arbitrarily well by simple functions of 
the form $f(x)=\sum_k x_k \chi_{B_k}(x)$ where $\cup_n B_k=\Rl$ is a 
collection of disjoint intervals and $x_k\in B_k$. Therefore 
for $\Omega_2$ in the domain of $a^n$
\ba \label{0.8}
<\Omega_1,a^n\Omega_2>&=&
\int\; x\;d\mu_{\Omega_1,a^{n-1}\Omega_2}(x)
\nonumber\\
&=& \lim \sum_{k_1} x_{k_1} <\Omega_1,E(B_{k_1})a^{n-1}\Omega_2>
\nonumber\\
&=& \lim \sum_{k_1} x_{k_1} <E(B_{k_1})\Omega_1,a^{n-1}\Omega_2>
\nonumber\\
&=& \lim \sum_{k_1} x_{k_1} \int x\;   
d\mu_{E(B_{k_1})\Omega_1,a^{n-2}\Omega_2}(x)
\nonumber\\
&=& \lim \sum_{k_1,k_2} x_{k_1} x_{k_2} 
<E(B_{k_2})E(B_{k_1})\Omega_1,a^{n-2}\Omega_2>
\nonumber\\
&=& ..
\nonumber\\
&=& \lim \sum_{k_1..k_n} x_{k_1}.. x_{k_n} 
<E(B_{k_1}\cap..\cap B_{k_n})\Omega_1,\Omega_2>
\nonumber\\
&=& \lim \sum_k\; x_k^n 
<\Omega_1,E(B_k)\Omega_2>
\nonumber\\
&=& \int\; x^n\; d\mu_{\Omega_1,\Omega_2}(x)
\ea
We conclude that for every measurable set $B$
\be \label{th0.9}
<\Omega_1,\chi_B(a)\Omega_2>=\int\; \chi_B(x)\; 
d\mu_{\Omega_1,\Omega_2}(x)=<\Omega_1,E(B)\Omega_2>
\ee
for all $\Omega_1,\Omega_2$ since $E(B)$ is a bounded operator. Thus
\be \label{0.10}
E(B)=\chi_B(a)
\ee
are the spectral projections associated with a self -- adjoint operator.
If we know the representation of $a$ on $\cal H$ then we have to 
approximate $\chi_B(a)$ by polynomials and then can construct the $E(B)$.
In particular we conclude that for every measurable function and 
$\Omega_2$ in the domain of $f(a)$  
\be \label{0.11}
<\Omega_1,f(a)\Omega_2>=\int\; f(x) \; d\mu_{\Omega_1,\Omega_2}(x)
\ee
since $f(E):=f(a)$ if $f(E)=\sum_n z_n E(B_n)$.
Formula (\ref{0.11}) is sometimes referred to as the functional calculus.
Combining (\ref{0.6}) and (\ref{0.10}) we have 
\be \label{0.11a}
\chi_{(-\infty,\lambda]}(a)=\theta(\lambda-a)=E((-\infty,\lambda])=
\int_{-\infty}^\lambda\;dE(x)=E(\lambda)-E(-\infty)=E(\lambda)
\ee
where the integration constant $E(-\infty)=0$ as follows from
the fact that $E((-\infty,-\infty))=E(\emptyset)=0$ by definition.\\
\\
Before we close this section we remark that the spectral theorem holds 
without making any separability assumptions, that is, it holds also when 
${\cal H}$ does not have a countable basis.

\subsection{Direct Integrals and Functional Models}
\label{s0.2}

\begin{Definition} \label{def0.2} ~~~\\
Let $(X,{\cal B},\mu)$ be a separable measure space such that $X$ is  
$\sigma-$finite with respect to $\mu$ 
and let 
$x\mapsto {\cal H}_x$ be an assignment
of separable Hilbert spaces such that the function $x\mapsto N(x)$, where 
$N(x)$
is the countable dimension of ${\cal H}_x$, is measurable.
It follows that the sets $X_N=\{x\in X;\;N(x)=N\}$,
where $N$ denotes any countable cardinality, are measurable. Since Hilbert 
spaces
whose dimensions have the same cardinality are unitarily equivalent we may
identify all the ${\cal H}_x,\;N(x)=N$ with a single ${\cal H}_N=\Cl^N$
with standard $l_2$ inner product. 
We now consider maps
\be \label{0.12}
\psi:\; X\to \prod_{x\in X} {\cal H}_x;\;\;
x\mapsto (\psi(x))_{x\in X}
\ee
subject to the following two constraints: \\
1. The maps $x\mapsto <\psi,\psi(x)>_{{\cal H}_N}$ are measurable for
all $x\in X_N$ and all $\psi \in {\cal H}_N$. \\
2. If
\be \label{0.13}
<\psi_1,\psi_2>:=\sum_N \int_{X_N}\; d\mu(x)
<\psi_1(x),\psi_2(x)>_{{\cal H}_N}
\ee
then $<\psi,\psi> <\infty$.\\
The completion of the space of maps (\ref{0.12}) in the inner product
(\ref{0.13}) is called the direct integral of the ${\cal H}_x$ with
respect to $\mu$ and one writes
\be \label{0.14}
{\cal H}^\oplus_{\mu,N}=\int_X^\oplus\;d\mu(x)\; {\cal H}_x,\;\;\;
<\xi_1,\xi_2>=
\int_X\;d\mu(x)\; <\xi_1(x),\xi_2(x)>_{{\cal H}_x}
\ee
\end{Definition}
The restriction to $\sigma-$finite measures is due to the fact that 
otherwise the Radon -- Nikodym theorem fails to hold \cite{Rudin}: If 
$\nu$ is a
finite positive measure absolutely continuous with respect to a 
finite positive 
measure on the same $(X,{\cal B})$ then there exists a $\mu-$a.e. positive 
$L_1(X,d\mu)$ function $\rho$ such that $\nu(B)=\int_B \rho d\mu$. If 
$\nu$
is only $\sigma-$finite and positive then $\rho$ is still positive 
$\mu-$a.e. 
but only 
measurable and not necessarily in $L_1(X,d\mu)$. If $\nu$ is not 
$\sigma-$finite 
then the Radon -- Nikodym theorem is false. In both cases one writes 
$\rho=d\nu/d\mu$. The Radon -- 
Nikodym theorem will prove crucial in our applications. The significance 
of separability of $\cal B$ is that such $\sigma-$algebras can be treated,
to arbitrary precision, as if they only had a 
countable number of elements. This implies that the space 
${\cal H}^\oplus_{\mu,N}$ is also separable: Consider 
functions $e_n$ with $e_n(x)\in {\cal H}_x$ and $e_n(x)=0$ for $n>N(x)$ 
such that $<e_n(x),e_m(x)>=\delta_{m,n}$ for $m,n\le N(x)$ and zero 
otherwise. For $x\in X_N$ the $e_n(x)=e^N_n$ are constant $\mu-a.e.$ and
provide an orthonormal basis on ${\cal H}_x={\cal H}_N$. Now fix any 
$f_0\in L_2(X,d\mu)$ such that $f_0\not=0$ $\mu-a.e.$. Then for 
every 
measurable $B$ from the assumed countable base the functions 
$e_{B,n}=f_0\chi_B e_n$ with 
$e_{B,n}(x)=f_0(x) \chi_B(x) e_n(x)$ are measurable and they obviously lie 
dense. Since the set of labels $(B,n)$ is countable, the Gram -- Schmidt
orthonormalization of the $e_{B,n}$ produces a countable basis for  
${\cal H}^\oplus_{\mu,N}$. Separability will also prove important for 
our applications. In what follows, we will always assume that $(X,{\cal 
B},\mu)$ is $\sigma-$finite and separable.
\begin{Definition} \label{def0.3} ~\\
Let ${\cal H}^\oplus_{\mu,N},\;{\cal H}^\oplus_{\mu,N'}$ be direct 
integral Hilbert spaces over 
$(X,{\cal B})$. Consider a family of fibre preserving, $\mu-$a.e. bounded 
operators 
$T(x)\in {\cal B}({\cal H}_x,{\cal H}'_x)$. The family is said to be 
measurable provided that the function $x\mapsto 
<\psi'(x),T(x)\psi(x)>_{{\cal H}'_x}$ is measurable for all $\psi\in {\cal H}_{\mu,N},\;\;\psi'\in {\cal H}_{\mu,N'}$. For a 
measurable family of fibre preserving operators one defines
\be \label{0.15}
<\psi',T\psi>_{{\cal H}^\oplus_{\mu,N'}}=\int_X\; d\mu(x)\;
<\psi'(x),T(x)\psi(x)>_{{\cal H}'_x}
\ee
In particular, if $N=N'$ $\mu-$a.e. and $T(x)$ is unitary then $T$ is 
called a measurable unitarity.
\end{Definition}
Direct integral Hilbert spaces carry the following natural fibre 
preserving, measurable operators:
Let for $B\in {\cal B}$ the operator $F(B)$ be defined 
by 
\be \label{0.15a}
(F(B)\psi)(x):=\chi_B(x) \psi(x)
\ee
Then it is easy to see that $F$ is a p.v.m. and the 
corresponding spectral measures are 
\be \label{0.16}
d\mu_\psi(x)=||\psi(x)||^2_{{\cal H}_x}\;d\mu(x)
\ee
so $\mu_\psi$ is absolutely continuous with respect to $\mu$. For any 
measurable scalar valued function $f$ the fibre preserving multiplication 
operator 
\be \label{0.17}
(Q_f\psi)(x):=f(x) \psi(x)
\ee
can be written in the spectral resolution form 
\be \label{0.18}
Q_f=\int_X \; f(x)\; dF(x)
\ee
~\\
The following theorem is the first step towards establishing a uniqueness 
result, up to unitary equivalence, of a direct integral representation 
subordinate to a self -- adjoint operator. We will denote by $[\mu]$ 
the equivalence class of all mutually absolutely continuous measures 
containing the representative $\mu$.
\begin{Theorem} \label{th0.2} ~\\
Suppose that two direct integral Hilbert spaces 
${\cal H}^\oplus_{\mu,N},\;{\cal H}^\oplus_{\mu',N'}$ over the same 
measurable space $(X,{\cal B})$ are given. \\
i)\\
If $[\mu]=[\mu']$ and $N=N'$ $\mu-$a.e., if $U$ is a fibre preserving 
measurable unitarity, then the operator 
$V:\;{\cal H}_{\mu,N}\to {\cal H}_{\mu',N'}$ defined by 
\be \label{0.19}
(V\psi)(x):=\sqrt{\frac{d\mu}{d\mu'}(x)}\; U(x)\;\psi(x)
\ee
is unitary and has the property $V F(B)=F'(B) V$ for all $B\in {\cal 
B}$.\\
ii)\\
If $V:\;{\cal H}_{\mu,N}\to {\cal H}_{\mu',N'}$ is a unitary operator 
satisfying $V F(B)=F'(B) V$ for all $B\in {\cal B}$ then $[\mu]=[\mu']$,
$N=N'$ $\mu$-a.e. and $V$ admits the presentation (\ref{0.19}).
\end{Theorem}
Proof of theorem \ref{0.19}:\\
i)\\
As defined, $V$ is certainly an isometry and since $[\mu]=[\mu']$ the 
function $d\mu'/d\mu$ is also positive $\mu'$-a.e. and clearly
$(d\mu'/d\mu) (d\mu/d\mu')=1$ $\mu$-a.e. Hence $V$ has an inverse, is thus 
unitary and the intertwining property $V F(B)=F'(B) V$ follows from the 
fibre preserving nature of all $V, F(B), F'(B)$ and because $F(B), F'(B)$ 
are just multiplication by scalars.\\
ii)\\
Let $f\in L_2(X,d\mu)$, $f\not=0$ $\mu-$a.e. and $<e_m(x),e_n(x)>_{{\cal 
H}_x}=\delta_{m,n}
\theta(N(x)-n)$ where $\theta(y)=1$ for $y\ge 0$ and $\theta(y)=0$ for 
$y<0$. Let $b_n(x):=f(x) e_n(x)$ then obviously 
$||b_n||_{{\cal H}_{\mu,N}}\le ||f||_{L_2(X,d\mu)}$ where equality is reached certainly for $n=1$ because $N(x)\ge 1$ $\mu-$a.e. Hence
$b_n\in {\cal H}_{\mu,N}$. Define $b'_n:=V b_n$. Then by unitarity and 
the intertwining property (unitary equivalence of the spectral 
projections)
\ba \label{0.20}
<b'_m,F'(B) b'_n>_{{\cal H}^\oplus_{\mu',N'}}   
& = &<b_m,F(B) b_n>_{{\cal H}^\oplus_{\mu,N}}  
\\
&\Rightarrow &
\int_B d\mu'(x) <b_m'(x),b_n'(x)>_{{\cal H}'_x}
=\int_B d\mu(x) <b_m(x),b_n(x)>_{{\cal H}_x}
\nonumber
\ea
Setting $m=n=1$ (\ref{0.20}) turns into 
\be \label{0.21}
\int_B d\mu'(x) ||b_1'(x)||^2_{{\cal H}'_x}
=\int_B d\mu(x) |f(x)|^2 
\ee
Since $B$ is arbitrary and $|f|^2$ positive $\mu-$a.e. while 
$||b_1'(x)||_{{\cal H}'_x}$ could vanish on sets of finite $\mu'-$measure 
we conclude that $\mu$ is absolutely continuous with respect to $\mu'$.
Interchanging the roles of $(\mu,N)$ and $(\mu',N')$ we see that actually
$[\mu]=[\mu']$. 

Set $\rho=d\mu/d\mu'$ and $m=n$ in (\ref{0.20}) then 
\be \label{0.22}
\int_B d\mu'(x) ||b_m'(x)||^2_{{\cal H}'_x}
=\int_B d\mu'(x) \rho(x) |f(x)|^2 \theta(N(x)-m) 
\ee
Since $B$ is arbitrary it follows that 
$||b_m'(x)||_{{\cal H}'_x}=\sqrt{\rho(x)} |f(x)| \theta(N(x)-m)$
$\mu-$a.e. while for $m\not=n$ the same argument 
leads to $<b'_m(x),b'_n(x)>=0$ $\mu-$a.e. Thus 
$<b'_m(x),b'_n(x)>=0$ for $m\not= n$ up to a zero measure set ${\cal 
N}_{m,n}$ and 
$<b'_m(x),b'_m(x)>=0$ up to a zero measure set ${\cal N}_{m,m}$ if 
$N(x)\ge 
m$. Let ${\cal N}:=\cup_{m,n} {\cal N}_{m,n}$ (countable collection).
Since $\mu$ is $\sigma-$additive we find $\mu({\cal N})=0$ so that 
our conclusions hold on a common set $X_0$ of full $\mu-$measure on which 
in particular $\rho>0$. It follows 
that $N'(x)\ge N(x)$ $\mu-$a.e. and interchanging $(N,\mu)$ and $(N.\mu')$
shows that $N=N'$ $\mu-$a.e.

On $X_0$ set $U(x)b_m(x):=\sqrt{\rho(x)}^{-1} b'_m(x)$ which defines a 
measurable isometry. Let $(V_1\psi)(x):=\sqrt{\rho(x)}U(x)\psi(x)$ then 
by definition of $b'_m$ and the intertwining property
\be \label{0.23}
(V_1 F(B) b_m)(x)=\sqrt{\rho(x)} U(x) \chi_B(x) b_m(x)=\chi_B(x) b'_m(x)
=(F'(B) V b_m)(x)=(V F(B) b_m)(x)
\ee 
Since the $F(B) b_m$ lie dense, $V=V_1$, so $V$ admits the presentation 
(\ref{0.19}).\\
$\Box$

The theorem reveals that direct integral Hilbert spaces on measurable 
spaces $(X,{\cal B})$ and their 
canonical p.v.m.'s  are uniquely characterized, up to unitary 
equivalence, by the type $[\mu]$ of the underlying measure $\mu$ and the 
multiplicity function
$N$. 

The next theorem characterizes all bounded operators that commute with the 
canonical p.v.m. of a direct integral Hilbert space.
\begin{Theorem} \label{th0.3} ~\\
Let ${\cal H}^\oplus_{\mu,N}$ be a direct integral Hilbert space.\\
i)\\
Suppose that $T$ is a 
measurable, fibre preserving, bounded -- operator -- valued function on 
${\cal H}^\oplus_{\mu,N}$ such that 
$\mu-\sup ||T(x)||_{{\cal H}_x}||<\infty$ (i.e. the operator norm in 
the fibres is uniformly bounded up to sets of measure zero). 
Then $T$ defined by $(T\psi)(x)=T(x)\psi(x)$ is a bounded operator on 
${\cal H}^\oplus_{\mu,N}$ which commutes with the canonical p.v.m.\\
ii)\\
If $T$ is a bounded operator on ${\cal H}^\oplus_{\mu,N}$ which commutes 
with the canonical p.v.m. then $T$ is a fibre preserving,
$\mu-$a.e. uniformly bounded operator. Moreover, the operator norm
coincides with the uniform fibre norm.
\end{Theorem}
Proof of theorem \ref{th0.3}:\\
i)\\
An elementary calculation shows that 
\be \label{0.24}
||T||_{{\cal H}^\oplus_{\mu,N}} \le \mu-\sup_{x\in X}  
||T(x)||_{{\cal H}_x}
\ee
so $T$ is bounded. That $[F(B),T]=0$ for all measurable $B$ is trivial.\\
ii)\\
Let $X_n=\{x\in X;\;N(x)=n\;\mu-$a.e.$\}=N^{-1}(n)$. These mutually 
disjoint sets 
are measurable due to measurability of $N$. If we set 
${\cal H}^\oplus_{\mu,n}:=\int_{X_n}\;d\mu(x)\; {\cal H}_x$ then clearly
\be \label{0.25}
{\cal H}^\oplus_{\mu,N}=\oplus_{n=1}^M 
{\cal H}^\oplus_{\mu,n}
\ee
where $M=\mu-\sup_{x\in X} N(x)$ is the maximal multiplicity.

Let $x\in X_m$ and $\psi\in {\cal H}^\oplus_{\mu,n}$ then by 
assumtion that $[F(B),T]=0$
\be \label{0.26}
(T \psi)(x)=
(T F(X_n) \psi)(x)=(F(X_n) T \psi)(x)=\chi_{X_n}(x) (T\psi)(x)=\delta_{mn}
(T \psi)(x)
\ee
Thus $T$ preserves all the ${\cal H}^\oplus_{\mu,n}$ and we may reduce the 
analysis to $x\in X_n$ so that $N(x)=n=const.$ and we may set ${\cal H}_x=
{\cal H}_n=const.$ on $X_n$.

Let ${\cal B}\ni B \subset X_n$ then for $\psi\in {\cal H}^\oplus_{\mu,n}$
due to boundedness
\ba \label{0.27} 
&& \int_{B} \;d\mu(x)\; ||(T\psi)(x)||^2_{{\cal H}_n}
=||F(B) T \psi||^2_{{\cal H}^\oplus_{\mu,N}}
=||T F(B) \psi||^2_{{\cal H}^\oplus_{\mu,N}}
\nonumber\\
& \le &  ||T||^2_{{\cal H}^\oplus_{\mu,N}}\;
||F(B) \psi||^2_{{\cal H}^\oplus_{\mu,N}}
=||T||^2_{{\cal H}^\oplus_{\mu,N}}\;
\int_{B} \;d\mu(x)\; ||\psi(x)||^2_{{\cal H}_n}
\ea
Since $B$ is arbitrary we conclude
\be \label{0.28}
||(T\psi)(x)||_{{\cal H}_n}
\le ||T||_{{\cal H}^\oplus_{\mu,N}}\;
||\psi(x)||_{{\cal H}_n}
\ee
$\mu-$a.e. Since $\psi$ could be supported on arbitrary measurable sets 
we conclude that $(T\psi)(x)\in {\cal H}_x$ must be fibre preserving. 
We may therefore define $T(x)\psi(x):=(T\psi)(x)$. Then (\ref{0.28}) 
automatically gives 
\be \label{0.28a}
||T(x)||_{{\cal H}_x}
\le ||T||_{{\cal H}^\oplus_{\mu,N}}\;
\ee
$\mu-$a.e. and $T(x)$ is uniformly bounded. Together with 
(\ref{0.24}) we obtain
\be \label{0.29}
||T||_{{\cal H}^\oplus_{\mu,N}} = \mu-\sup_{x\in X}  
||T(x)||_{{\cal H}_x}
\ee
$\Box$\\
We see that bounded operators commuting with the canonical p.v.m.
are precisely the fibre preserving bounded operator valued
functions on the direct integral. Such bounded operators are called 
decomposable. Since for fibre preserving bounded operators 
no domain questions arise, the operations 
of scalar multiplication, addition, multiplication and taking adjoints 
can be done fibre -- wise and we see that bounded, self -- adjoint, 
unitary, 
normal and projection operators in the commutant of the canonical
p.v.m. are precisely the fibre preserving operators which are 
self -- adjoint, unitary, normal and projection operators in every fibre
$\mu-$a.e. Finally, if $T$ is a bounded operator which commutes with every 
decomposable operator then it must be itself decomposble because all the 
$F(B)$ are decomposble. Therefore its fibre component must commute with 
every bounded operator on ${\cal H}_x$ and thus $T(x)=q(x) 1_{{\cal H}_x}$ 
is multiplication operator by a scalars 
on each fibre.

\subsection{Direct Integral Representation of Projection Valued Measures}
\label{s0.3}

Let $\cal H$ be a separable Hilbert space and $E$ a p.v.m. on 
a 
measurable space $(X,{\cal B})$. Choose a unit vector 
$\Omega_1\in {\cal H}$ and let ${\cal H}_{\Omega_1}$ be the closure 
of the vector space of vectors of the form 
$[\sum_{k=1}^K z_k E(B_k)]\Omega_1$ where $z_k\in \Cl,\;K<\infty,\;B_k\in 
{\cal B}$. If ${\cal H}_{\Omega_1}\not={\cal H}$ choose $\Omega_2\in 
{\cal H}_{\Omega_1}^\perp$ and construct ${\cal H}_{\Omega_2}$. Clearly
${\cal H}_{\Omega_1}\perp {\cal H}_{\Omega_2}$. Suppose that mutually 
orthogonal 
${\cal H}_{\Omega_1},..,{\cal H}_{\Omega_n}$ have already been constructed   
but that 
${\cal H}_{\Omega_1}\oplus ..\oplus {\cal H}_{\Omega_n}\not={\cal H}$.
Then choose $\Omega_{n+1}$ in the orthogonal complement and construct 
${\cal H}_{\Omega_{n+1}}$ which is orthogonal to all the other spaces.
The procedure must come to an end after at most a countable number $M$ of 
steps because the $\Omega_n$ form a countable orthogonal system and 
${\cal H}$ has a countable basis.

We consider the associated spectral measures 
$\mu_{\Omega_n}(B)=<\Omega_n,E(B)\Omega_n>_{{\cal H}}$ and the total 
measure 
\be \label{0.30}
\mu_\Omega(B):=\sum_{n=1}^M c_n \mu_{\Omega_n}(B)
\ee
where $\Omega:=\sum_{n=1}^M c_n^{1/2} \Omega_n$ and $c_n>0,\;\sum_{n=1}^M
c_n=1$ are any positive constants. It is often convenient to choose
$c_n=2^{-n}/\sum_{n=1}^M 2^{-n}$.

Notice that $\mu_\Omega(B)=<\Omega, E(B)\Omega>$ and that all measures 
are probability measures. The measure 
$\mu_\Omega$ has the following maximality feature:
\begin{Lemma} \label{la0.1} ~\\
For any $\Psi\in {\cal H}$ the associated spectral measure 
$\mu_\Psi(B)=<\Psi,E(B)\Psi>$ is absolutely continuous with respect to 
$\mu_\Omega$.
\end{Lemma}
Proof of lemma \ref{la0.1}:\\
A dense set of vectors in $\cal H$ is of the form $\Psi=\sum_n \Psi_n$ 
with $\Psi_n=\sum_{k=1}^\infty z^n_k E(B_k^n) \Omega_n$ and $z_k^n=0$
for all but finitely many $k$.  
The Hilbert space ${\cal H}_{\Omega_n}$ is unitarily equivalent to 
$L_2(X,d\mu_{\Omega_n})$ via $\Psi_n\mapsto
 \sum_{k=1}^\infty z^n_k \chi_{B_k^n}$. It follows that 
\be \label{0.31} 
\mu_{\Psi_n}(B)=\int_B\;d\mu_{\Omega_n}(x)\;|\Psi_n(x)|^2
\ee
hence $\mu_{\Psi_n}$ is absolutely continuous with respect to 
$\mu_{\Omega_n}$. Since every $\mu_{\Omega_n}$ is absolutely continuous 
with respect to $\mu_{\Omega}$, the claim follows.\\
$\Box$\\
Thus, while the choice of $\Omega_n,\;c_n$ and thus $\Omega$ 
is not unique, the type of $\mu_{\Omega}$ is unique.
\begin{Definition} \label{def0.4} ~\\
Let $E$ be a p.v.m.. The type $[E]$ of $E$ is given by
$[\mu_\Omega]$ where $\Omega$ is any vector satisfying the maximality 
criterion of lemma \ref{la0.1} (which was shown to exist).
\end{Definition}
Notice that this would not hold if $\cal H$ is not separable. In the 
non -- separable case it may take an uncountable collection of $\Omega_n$
in order to decompose ${\cal H}$ as above. Then $\mu=\sum_n c_n 
\mu_{\Omega_n}$ may still be formed but in order to be well -- defined 
(not identical to the measurue which is infinite a.e.) we generically 
would need to set all but a countable number of the $c_n$ equal to zero. 
But then for $c_n=0$ we do not have that $\mu_{\Omega_n}$ is absolutely 
continuous with respect to $\mu_\Omega$. Hence separability is essential 
in order that the type $[E]$ be well -- defined.

Consider any collection $\Omega_n$ such that $\Omega$ leads to maximal 
type $[E]$. Since all the measures $\mu_{\Omega_n}$ and $\mu_\Omega$
are actually finite measures, the Radon -- Nikodym derivatives
$\rho_n(x):=d\mu_{\Omega_n}(x)/d\mu_\Omega(x)$ exist and are non -- 
negative 
$L_1(X,d\mu)$ functions. As such they are only defined $\mu-$a.e. 
but let us pick any representative. 
Let $S_n:=\{x\in X;\;\rho_n(x)>0\}$ be the 
support of these functions. These sets are measurable because the 
functions $\rho_n$ are measurable (namely $S_n$ is the support of 
$\mu_{\Omega_n}$ which is $\mu_{\Omega_n}-$measurable and thus 
$\mu_\Omega$ measurable). Given $x\in X$ we define $N_E(x)=n$ if there are 
precisely $n$ integers $k_1(x)<..<k_n(x)$ such that $x\in S_k,\;
k\in \{k_1,..,k_n\}$. The function $N_E:\;X\to \Nl$ is measurable because 
$\Nl$ carries the discrete topology (all sets are open, in particular the 
one point sets $\{n\}$) and 
\be \label{0.32}
X_n:=N_E^{-1}(\{n\})=X'_n-X'_{n+1} \mbox{ 
where } X'_n:=\cup_{k_1<..<k_n}\;\cap_{l=1}^n S_{k_l}
\ee
is the set of points which are in at least $n$ of the supports of the 
$\rho_n$ (notice that $X'_{n+1}\subset X'_n$). Thus,
$X_n$ is the difference of a countable union of measurable sets, hence 
it is measurable for every $n\in \Nl$, thus $N_E$ is a measurable 
function.\\ 
Furthermore, the functions $k_l(x),\;l=1,..,N(x)$ are 
measurable. To see this, consider the function $K:\;X\mapsto P(\Nl);\;
x\mapsto\{k_1(x),..,k_{N(x)}(x)\}$ where $P(S)$ denotes the power set
(set of all subsets of) the set $S$. We have $K^{-1}(\{k_1,..,k_N\})=
S_{k_1}\cap..\cap S_{k_N}-X'_{N+1}\subset X_N$ so $M$ is clearly 
measurable because $P(\Nl)$
carries the discrete topology. Next let $n:\;P(\Nl)\to \Nl\cup\{0\};\;
\{k_1,..k_N\}\mapsto k_n$ if $n\le N$ with the convention that 
$k_1<..k_N$ and otherwise $\{k_1,..k_N\}\mapsto 0$. The function $n$ is 
continuous because it maps between discrete topologies. Now 
$k_n(x)=(n\circ K)(x)$ is the composition of a measurable with a 
continuous function, hence it is measurable. 

As an example choose ${\cal H}={\cal H}_{\mu,N}$ and $E=F$ the canonical 
p.v.m. Let us show that  
$[F]=[\mu]$ and $N_F=N$. To see this,
recall that the functions $e_{B,n}$ with $e_{B,n}(x)=f(x) \chi_B(x) 
e_n(x)$ where $f\in L_2(X,d\mu),\;f\not=0$ $\mu-$a.e. and 
$<e_m(x),e_n(x)>_{{\cal H}_x}=\delta_{m,n} \theta(N(x)-n)$ are dense in
${\cal H}^\oplus_{\mu,N}$. We may therefore choose $\Omega_n=f e_n$. 
We calculate 
\be \label{0.33}
d\mu_{\Omega_n}(x)/d\mu(x)=|f(x)|^2 ||e_n(x)||^2_{{\cal H}_x}
=|f(x)|^2 \theta(N(x)-n)
\ee
Therefore, choosing not to normalize the $||\Omega_n||\le 
1$ 
\be \label{0.34}
d\mu_\Omega(x)\propto \sum_n 2^{-n} d\mu_{\Omega_n}(x)=|f(x)|^2 d\mu(x)
\sum_{n=1}^{N(x)} 2^{-n}=[1-2^{-N(x)}] |f(x)|^2 d\mu(x)
\ee
Hence, $d\mu_\Omega$ and $|f|^2 d\mu$ are equivalent measures because 
$N(x)\ge 1$ $\mu-$a.e. by convention (otherwise restrict $X$). 
This reveals $[F]=[\mu]$. It follows 
that $S_n=\{x\in X;\theta(N(x)-n)>0\}$ and $N_F(x)=n$ if $x$ lies 
precisely in $n$ of the $S_k$, say $k_1<..<k_n$ which is 
precisely the case for $N(x)=k_n$. However, if $k_n>n$ then this is 
the case for all $k=1,..,k_n$. Therefore $k_l=l,\;l=1,..,n$ so 
that $N(x)=n$. Thus $N_F(x)=N(x)$ $\mu-$a.e.  

The worry is now that the function $N_E$ depends not only on $E$ but also 
on the choice of the $\Omega_n$. This is excluded by the following theorem
which proves more: $N_E$, as a measurable function, depends only on $E$ 
and, moreover, any p.v.m. $E$ is completely characterized 
by the type $[E]$ and the muliplicity function $N_E$ up to unitary 
equivalence. 
\begin{Theorem} \label{th0.5} ~\\
Let $E$ be a p.v.m. on a Hilbert space $\cal H$ and 
${\cal H}_{\mu,N}$ a direct integral Hilbert space together with its 
natural p.v.m. $F$. Of course, both $E$ and $\mu$ are based 
on the same measurable space $(X,{\cal B})$.\\
i)\\
If $[E]=[\mu]$ and $N_E(x)=N(x)$ $\mu-$a.e. then there is a unitary 
operator $V:\;{\cal H}\to {\cal H}_{\mu,N}$ such that 
$V E(B)=F(B) V$ for all measurable $B\in {\cal B}$.\\
ii)\\
a) $N_E$ only depends on $E$ and not on the concrete choice of the 
$\Omega_n,\;c_n$ that lead to $N_E$ as above.\\
b) The spectral type $[E]$ and the multiplicity function $N_E$ are 
unitary invariants of $E$.\\
iii)\\
The data $([E],N)$ determine $E$ up to unitary equivalence.
\end{Theorem}
Proof of theorem \ref{th0.5}:\\
Without loss of generality we may assume that the measure $\mu$ underlying
${\cal H}_{\mu,N}$ is finite because by using an $f\in L_2(X,d\mu)$ with 
$f\not=0$ $\mu-$a.e. we can switch to the finite measure $d\mu'=|f|^2 
d\mu$
which gives rise to $[\mu']=[\mu]$ and $N=N'$ $\mu-$a.e., hence by 
theorem \ref{th0.2} the corresponding direct integral Hilbert spaces and 
the canonical p.v.m. are unitarily equivalent.\\
i)\\
By assumption $[E]=[\mu]$ and $N_E(x)=N(x)$ $\mu-$a.e. Given $x\in X$ 
let $M(x)=\{n_1(x)<..<n_{N(x)}\}$ be the set of indices $n$ such that 
$x\in S_{\Omega_n}$. Any vector $\Psi\in {\cal H}$ can be written in the 
form $\Psi=\sum_{n=1}^M \Psi_n(E) \Omega_n$ for measurable $\Psi_n$ and has 
the norm 
\ba \label{0.35} 
||\Psi||^2_{{\cal H}} 
&=& \sum_{n=1}^M \int_X \;d\mu_\Omega(x)\; \rho_n(x) |\Psi_n(x)|^2
\nonumber\\
&=&  \int_X \;d\mu_\Omega(x)\; \sum_{n=1}^M \rho_n(x) |\Psi_n(x)|^2
\nonumber\\
&=& \sum_{m=1}^M \int_{X_m} \;d\mu_\Omega(x)\; \sum_{n=1}^M \rho_n(x) 
|\Psi_n(x)|^2
\nonumber\\
&=& \sum_{m=1}^\infty \int_{X_m} \;d\mu_\Omega(x)\; \sum_{n\in M(x)} 
\rho_n(x) 
|\Psi_n(x)|^2
\nonumber\\
&=& \sum_{m=1}^\infty \int_{X_m} \;d\mu_\Omega(x)\; \sum_{n=1}^m 
\rho_{k_n(x)}(x) |\Psi_{k_n(x)}(x)|^2
\ea
In the second step we have used the fact that $L_2(X,d\mu;l_2^M)\cong
l_2^M(L_2(X,d\mu))$, i.e. the Hilbert space of square integrable vector 
valued functions with values in the Hilbert space $l_2^M$ of square 
summable sequences (with label set $n=1,..,M\le \infty$) is isometrically 
isomorphic to the Hilbert space of 
square summable sequences of square integrable functions. This allowed us 
to interchange the sum and the integral (alternatively, use the Lebesgue 
dominated convergence theorem). In the third step we have decomposed $X$ 
into the $X_m=N^{-1}(m)$, recall (\ref{0.32}), and then could restrict the 
sum over $n$ to the $k_1(x),..,k_m(x)$.    

For $x\in X_m$ consider an orthonormal basis $e^{(m)}_n,\;n=1,..,m$ of the 
Hilbert space $\Cl^m$ equipped with the standard inner product. We extend 
this to vector valued functions $e_n(x)$ where 
\be \label{0.36}
e_n(x)=\sum_{m=1}^\infty \chi_{X_m}(x) \theta(N(x)-n) e^{(m)}_n 
\ee
and set ${\cal H}_x=\Cl^m$ if $x\in X_m$. Let 
\be \label{0.37}
\psi(x):=\sum_{n=1}^{N(x)} \sqrt{\rho_{k_n(x)}(x)} \Psi_{k_n(x)}(x) 
e_n(x)
\ee
Then (\ref{0.35}) shows that the map 
$V_1:\;{\cal H}\to {\cal H}_{\mu_\Omega,N};\;\Psi\mapsto \psi$ is an 
isometry. 

To show that it is unitary, we must show that its image is dense. Let 
$B\in {\cal B}$ and fix $n_0\in \Nl$. Consider 
\be \label{0.38}
\Psi^{B,n_0}_n(x):=\chi_B(x) \theta(N(x)-n_0) 
\delta_{n,k_{n_0}(x)}/\sqrt{\rho_{k_{n_0}(x)}(x)}
\ee
The function (\ref{0.38}) is measurable: $\chi_B$ is obviously measurable,
$x\mapsto \theta(N(x)-n_0)$ is the composition of the measurable map 
$N$ and the continuous map $k\mapsto \theta(k-n_0)$ on $\Nl$ (discrete 
topology), 
$x\mapsto \delta_{n,k_{n_0}(x)}$ is the composition of the measurable map
$k_{n_0}$ (see above) and the continuous map $k\mapsto \delta_{n,k}$ on 
$\Nl$, $\rho_{k_{n_0}(x)}(x)=\sum_{m=1}^M \delta_{m,k_{n_0}(x)} \rho_m(x)$ and 
now we just need to use the fact that taking sums and products 
are continuous maps $\Rl\times \Rl\to \Rl$ (product topology) and taking 
square roots is a continuous map on $\Rl_+$ to see that (\ref{0.38}) is 
measurable. It is easy to check that $V_1\Psi^{B,n_0}=F(B) e_{n_0}$.
Since the functions $F(B) e_n$ are dense in ${\cal H}_{\mu_\Omega,N}$ it 
follows that $V_1$ is unitary. Moreover, it is trivial to see that 
$V_1 E(B)=F(B) V_1$.

Now consider the given direct integral Hilbert space ${\cal H}_{\mu,N}$. 
By assumption we have $[\mu_\Omega]=[\mu]$. Hence, choosing 
$U(x)=\sqrt{d\mu_\Omega/d\mu}(x)\; 1_{{\cal H}_x}$ in theorem \ref{th0.2} 
i) as the measurable 
unitarity 
we conclude that there exists a unitary operator $V_2:\;{\cal 
H}_{\mu_\Omega,N} \to {\cal H}_{\mu,N}$ satisfying $V_2 F(B)=F'(B) 
V_2$ where we have also denoted by $F$ the canonical p.v.m. on ${\cal 
H}_{\mu,N}$ therby slightly abusing the notation.
Thus, $V=V_2 V_1:\;{\cal H}\to {\cal H}_{\mu,N}$ is the searched for 
unitarity satisfying $V E(B)=F'(B) V$ for all $B\in {\cal B}$.\\
ii)\\
a) Consider two sets of vectors and constants
$(\Omega_n,\;c_n)_{n=1}^M,\;(\Omega'_n,\; c'_n)_{n=1}^{M'}$ leading to two 
multiplicity 
functions $N_E,N'_E$ but of course to equivalent measures $\mu_E,\;\mu'_E$
of type
$[E]=[\mu_E]=[\mu'_E]$. Using theorem \ref{th0.5} i) we find 
unitary operators $V:\;{\cal H}\to {\cal H}_{\mu_E,N_E}$ 
and $V':\;{\cal H}\to {\cal H}_{\mu'_E,N'_E}$ satisfying 
$V E(B)=F(B) V$ and  $V' E(B)=F'(B) V'$ for all $B\in {\cal 
B}$ respectively. It follows that $\tilde{V}=V' V^{-1}:\;
{\cal H}_{\mu_E,N_E} \to {\cal H}_{\mu_E,N_E}$ is a unitary operator
satisfying $\tilde{V} F(B)=F'(B) \tilde{V}$. Hence, by theorem \ref{th0.2} 
ii) we find that $N_E(x)=N'_E(x)$ $\mu_E-$a.e.\\
b)\\
Suppose we are given two p.v.m.'s $E_j$ on Hilbert spaces
${\cal H}_j,\;j=1,2$ which are unitarily equivalent, that is, there is a 
unitarity $V:\;{\cal H}_1\to {\cal H}_2$ satisfying $V E_1(B)=E_2(B) V$ 
for all $B\in {\cal B}$. By theorem \ref{th0.5} i) we find unitary maps
$V_j:\;{\cal H}_j\to {\cal H}_{\mu_{E_j},N_{E_j}}$ satisfying 
$V_j E_j(B)=F_j(B) V_j$. Thus the unitary operator 
$\tilde{V}:=V_2 V V_1^{-1}:\;{\cal H}_{\mu_{E_1},N_{E_1}} 
\to {\cal H}_{\mu_{E_2},N_{E_2}}$ satisfies $\tilde{V} 
F_1(B)=F_2(B)\tilde{V}$. Thus, by theorem \ref{th0.2} ii) we have 
$[\mu_{E_1}]=[\mu_{E_2}]$ and $N_{E_1}=N_{E_2}$ $\mu_{E_1}-$a.e.\\
iii)\\
Given $([E],N)$ choose representatives $E_1,E_2$ with 
$[E_j]=[E]$ and $N_{E_j}=N$ a.e. The corresponding unitarily 
equivalent models $F_j$ on 
${\cal H}_{\mu_{E_j},N_{E_j}}$ satisfy $[\mu_{E_1}]=[\mu_{E_2}],\;
N_{E_1}=N_{E_2}$ a.e. by assumption, hence by theorem \ref{th0.2} i)
the models are unitarily equivalent and so are $E_1, E_2$. \\
$\Box$\\
\\
Remarks:\\
1.\\
Given a self -- adjoint operator $a$ represented on a Hilbert space ${\cal 
H}$ it is often easier to choose the $\Omega_n$ and to determine the 
Hilbert spaces ${\cal 
H}_{\Omega_n}$ as follows: ${\cal H}$ has a dense set of 
$C^\infty-$vectors $\Omega$ for $a$, that is, vectors which are in the 
domain of 
$a^n$ for all $n=0,1,2,..$, see e.g. \cite{RS}. To see that the closure of 
the span of the $a^n\Omega$ coincides with the span of the 
$E(B)\Omega,\;B\in {\cal B}$ we notice that every measurable function 
on $\Rl$ can be approximated $\mu-$a.e. to arbitrary precision by smooth 
functions 
of rapid decrease which in turn have convergent Taylor expansions. 
The same is true for the linear span of the $\chi_B$. Thus, given that 
mutually orthogonal ${\cal H}_{\Omega_n}$ with $\Omega_n$ 
$C^\infty-$vectors for $a$ have already been constructed, we restrict 
$a$ to the orthogonal complement of those spaces and simply choose a
$C^\infty-$vector in that space.\\
2.\\
It is now evident why the function $N(x)$ is called multiplicity function:
In each fibre ${\cal H}_x$ the operator $a$ acts by multiplication by $x$
as follows from the spectral theorem, we also formally derive this in 
section \ref{0.5}. Since $N(x)=\dim({\cal H}_x)$ this means that the 
(generalized) eigenvalue $x$ has multiplicity $N(x)$. Furthermore, on 
${\cal H}^N_\mu:=\int_{X_N}^\oplus d\mu(x) {\cal H}_x$ the operator $a$ 
has 
constant multiplicity $N$ and we have ${\cal H}\cong \oplus_{N=1}^M
{\cal H}^N_\mu$. It follows that $a$ is multiplicity free ($N\equiv 1$) if 
and only if there is a cyclic vector for $a$. 

\subsection{Direct Integral Representations and Spectral Types}
\label{s0.4}

We will need this section in order to properly deal with constraint 
quantization.\\
\\
Given the measures $\mu_n:=\mu_{\Omega_n},\mu:=\mu_{\Omega}=\sum_n c_n 
\mu_n$ denote by 
$P_n,P$ respectively their pure point sets, that is, the set of points 
$p\in X$ such that $\mu_n(\{p\})>0$ and $\mu(\{p\})>0$ respectively.
Obviously $P=\cup_n P_n$. Since the measures $\mu_n,\mu$ are probability 
measures the sets $P_n,P$ must have countable cardinality (this is not 
necessarily the case when ${\cal H}$ is not separable in which case the 
label set of the $n$ may have uncountable cardinality). We define for
$B\in {\cal B}$ the pure point measures
\be \label{0.40}
\mu^{pp}_n(B):=\mu_n(P_n\cap B),\;\;
\mu^{pp}(B):=\mu(P\cap B)
\ee
Then $\mu^c_n:=\mu_n-\mu^{pp}_n,\;\mu^c:=\mu-\mu^{pp}$ are positive 
measures with the property that their sets of pure points is empty because
e.g. $\mu^c(B)=\mu(B-P)$. They are called continuous for that reason. 
Thus the measures allow for a unique decomposition, e.g. 
$\mu=\mu^{pp}+\mu^c$
into their pure point and continuous part respectively. In the case of 
interest, $X=\sigma(a)$ is a subset of $\Rl$ which carries the natural 
Borel $\sigma-$algebra ${\cal B}_{Borel}$. The measurable space 
$(\Rl,{\cal B}_{Borel})$ carries the natural 
$\sigma-$finite Lebesgue measure $d\mu_L(x)=dx$ and the Lebesgue 
decomposition theorem tells us that every 
measure $\mu$ on $(\Rl,{\cal 
B}_{Borel})$ can be uniquely
decomposed into $\mu=\mu^{ac}+\mu^s$ where $\mu^{ac},\;\mu^s$ 
respectively is 
absolutely continuous and singular with respect to $\mu_L$ 
respectively (that is, $\mu_L(B)=0$ implies $\mu^{ac}(B)=0$ while 
there exists a measurable set $S$ such that $\mu^s(S)=0$ and 
$\mu_L(\Rl-S)=0$). Since $\mu_L$ has no pure points we may apply the above 
observation to split $\mu^s$ further as $\mu^{pp}+\mu^{cs}$ where 
$\mu^{cs}$ is the continuous singular part of $\mu$. 

Coming back to our concrete direct integral construction $\mu=\sum_n c_n 
\mu_n$ with 
$c_n>0,\;\sum_n c_n=1$ we may decompose each of the $\mu_n$ and $\mu$ 
independently into the parts $\ast=pp,ac,cs$. 
We derive two simple results:
\begin{Lemma} \label{la0.2} ~\\
We always have
$\mu^\ast=\sum_n c_n\mu_n^\ast$ for $\ast=pp,ac,cs$.  
\end{Lemma}
Proof of lemma \ref{la0.2}:\\
By definition, using 
$P_n\subset P$
\be \label{0.41}
\mu^{pp}(B)=\mu(B\cap P)=
\sum_n c_n \mu_n(B\cap P)
=\sum_n c_n [\mu_n(B\cap P_n)+\mu_n(B\cap(P-P_n))]
=\sum_n c_n \mu_n^{pp}(B)
\ee
because $B\cap (P-P_n)$ is a discrete set containing no pure points of 
$\mu_n$. It follows that $\mu^c=\mu^{ac}+\mu^{cs}=\sum_n 
c_n[\mu_n^{ac}+\mu_n^{cs}]=\sum_n c_n \mu^c_n$. 
Next, let $S_n,\;n=1,..,M$ be such that 
$\mu^{cs}_n(S_n)=0$ and $\mu_L(\Rl-S_n)=0$ and let $S_0$ be such that 
$\mu^{cs}(S_0)=0$ and $\mu_L(\Rl-S_0)=0$. 
Define $S=\cap_{n=0}^M S_n$.
Then $\mu_L(\Rl-S)=\mu_L(\cup_{n=0}^M (\Rl-S_n))\le \sum_{n=0}^M
\mu_L(\Rl-S_n)=0$
by $\sigma-$additivity. This implies $\mu^{ac}(\Rl-S)=\mu_n^{ac}(\Rl-S)=0$
for all $n$ due to absolute continuity. Moreover, 
$\mu^{cs}(S)=\mu_n^{cs}(S)=0$ for 
all $n=1,2,..$ since $S\subset S_n$ for all $n=0,1,2,..$. Thus, if
$B\subset \Rl-S$ then $\mu^c(B)=\mu^{cs}(B)=\sum_{n=1}^M c_n 
\mu_n^{cs}(B)$ and if $B\subset S$ then $\mu^{cs}(B)=\sum_{n=1}^\infty 
c_n \mu_n^{cs}(B)=0$ anyway. Thus 
$\mu^{cs}(B)=\sum_{n=1}^\infty \mu_n^{cs}(B)$ for all $B$. It follows that   $\mu^{ac}(B)=\sum_{n=1}^\infty \mu_n^{ac}(B)$ for all $B$ as well.\\
$\Box$\\
\begin{Lemma} \label{la0.3}  ~\\
Given two unitarily equivalent direct integral representations ${\cal H}_{\mu,N},\;{\cal H}_{\mu',N'}$ of $a$ so that necessarily $[\mu]=[\mu']$
and $N=N'$ $\mu-$a.e. we always have 
$[\mu^\ast]=[\mu^{\prime \ast}]$ for $\ast=pp,ac,cs$. 
\end{Lemma}
Proof of lemma \ref{la0.3}:\\
Let $P,P'$ be the pure points of $\mu,\mu'$ respectively.
Then $\mu(P'-P)=0$ implies $\mu'(P'-P)=0$ by absolute continuity, hence 
$P'\subset P$. Likewise,   
$\mu'(P-P')=0$ implies $\mu(P-P')=0$ by absolute continuity, hence 
$P\subset P'$. Thus $P=P'$ and so $[\mu^{pp}]=[\mu^{\prime pp}]$.

Let $S,S'$ be such that 
$\mu_L(\Rl-S)=\mu_L(\Rl-S')=\mu^{cs}(S)=\mu^{\prime cs}(S')=0$.
Then also $\mu_L(\Rl-\tilde{S})=\mu^{cs}(\tilde{S})=\mu^{\prime 
cs}(\tilde{S})=0$ where $\tilde{S}=S\cap S'$. 

Suppose 
$B\subset \Rl-(\tilde{S}\cup P)$. Then $\mu(B)=\mu^{cs}(B)=0$ if and only 
if 
$\mu^{\prime}(B)=\mu^{\prime cs}(B)=0$ by absolute continuity. If 
$B\subset \tilde{S}\cup P$ then anyway $\mu^{cs}(B)=\mu^{\prime cs}(B)=0$,
hence $\mu^{cs}(B)=0 \; \Leftrightarrow \mu^{\prime cs}(B)=0$ for all 
$B\in {\cal B}$, that is, $[\mu^{cs}]=[\mu^{\prime cs}]$.

Suppose 
$B\subset \tilde{S}-P$. Then $\mu(B)=\mu^{ac}(B)=0$ if and only 
if 
$\mu^{\prime}(B)=\mu^{\prime ac}(B)=0$ by absolute continuity. If 
$B\subset \Rl-(\tilde{S}-P)\subset (\Rl-\tilde{S})\cup P$ then anyway 
$\mu^{ac}(B)=\mu^{\prime ac}(B)=0$,
hence $\mu^{ac}(B)=0 \; \Leftrightarrow \mu^{\prime ac}(B)=0$ for all 
$B\in {\cal B}$, that is, $[\mu^{ac}]=[\mu^{\prime ac}]$.\\
$\Box$
\begin{Lemma} \label{la0.4} ~\\
Let ${\cal H}^\ast=\{\Psi\in {\cal H};\;\mu_\Psi=\mu_\Psi^\ast\}$ where 
$\ast\in\{pp,ac,cs\}$ and $\mu_\Psi(.)=<\Psi,E(.)\Psi>$ denotes the 
spectral measure of $\Psi$. Then ${\cal H}={\cal H}^{pp}\oplus {\cal 
H}^{ac}\oplus {\cal H}^{cs}$. Moreover, each space ${\cal H}^\ast$ is 
invariant under the $E(B),\;B\in {\cal B}$.
\end{Lemma}
Proof of lemma \ref{la0.4}:\\
Let $\Omega_n$ be a cyclic system such that the 
${\cal H}_n=\overline{\mbox{span}\{E(B)\Omega_n;\;B\in {\cal B}\}}$
are mutually orthogonal and ${\cal H}=\oplus_n {\cal H}_n$. Let $\mu_n$
be the spectral measure of $\Omega_n$ and let the $\Omega_n$ be normalized
such that $\mu=\sum_n \mu_n$ is a probability measure. Any $\Psi\in {\cal 
H}$ is of the form $\Psi=\sum_n \Psi_n(E)\Omega_n$ with measurable 
functions $\Psi_n$ and thus 
\be \label{0.41a}
\mu_\Psi(B)=\sum_n \int_B \; d\mu_n \;|\Psi_n|^2
=\int_B \; d\mu \;[\sum_n \rho_n |\Psi_n|^2]
\ee
i.e. $d\mu_\Psi=|\Psi|^2 d\mu$ with $\rho_n=d\mu_n/d\mu$ and 
$|\Psi|^2:=\sum_n \rho_n |\Psi_n|^2$. Writing 
$\mu=\mu^{pp}+\mu^{ac}+\mu^{cs}$, let $P$ be the pure points of 
$\mu$ and $\mu_L(\Rl-S)=\mu^{cs}(S)=0$. We may assume without loss of 
generality that $S\cap P=\emptyset$. We may write (\ref{0.41a}) as
\be \label{0.41b}
\mu_\Psi(B)
=\int_B \; d\mu^{pp} \;|\Psi|^2
+\int_B \; d\mu^{ac} \;|\Psi|^2
+\int_B \; d\mu^{cs} \;|\Psi|^2
\ee
which gives rise to a map $V:\;{\cal H}=L_2(\Rl,d\mu)\to  
L_2(\Rl,d\mu^{pp})\oplus L_2(\Rl,d\mu^{ac})\oplus L_2(\Rl,d\mu^{cs})$
defined by $\Psi\mapsto (E(P)\Psi,E(S)\Psi,E(\Rl-(S\cup P))\Psi)$. Recall 
from 
the previous section that the type of the measure $\mu$ is uniquely 
determined and by lemma \ref{la0.3} the type of the measures $\mu^\ast$ 
is also uniquely determined, hence the sets $P,S$ are also determined
uniquely $\mu-$a.e. The map $V$ is an isometry by (\ref{0.41b}) 
and it is invertible because 
$E(P)+E(S)+E(\Rl-(P\cup S))=1_{{\cal H}}$, hence it is unitary.

We must show that ${\cal H}^\ast=L_2(\Rl,d\mu^\ast)$ for $\ast=pp,ac,cs$
or, in other words, that 
${\cal H}^{pp}=E(P){\cal H},\;
{\cal H}^{ac}=E(S){\cal H},\;
{\cal H}^{cs}=E(\Rl-(S\cup P)){\cal H}$.
Since $d\mu_{E(B)\Psi}=\chi_B d\mu_\Psi=\chi_B |\Psi|^2 d\mu$ it is clear 
that $d\mu_{E(B^\ast)\Psi}=|\Psi|^2 d\mu^\ast$ where 
$B^\ast=P,S,[\Rl-(S\cup P)]$ for $\ast=pp,ac,cs$. Hence 
$E(B^\ast){\cal H}\subset {\cal H}^\ast$. Conversely, since 
$d\mu_\Psi=|\Psi|^2 d\mu$, it follows that $\mu_\Psi$ is absolutely 
continuous with respect to $\mu$. Thus, by the method of proof of lemma
\ref{la0.3} it follows that $\mu_\Psi^\ast$ is absolutely continuous with 
respect to $\mu^\ast$. Hence, if $\Psi\in {\cal H}^\ast$, i.e. $\mu_\Psi=
\mu_\Psi^\ast$ then $d\mu_\Psi=|\Psi|^2 d\mu^\ast$ so 
$\Psi\in E(B^\ast){\cal H}$. 

That $E(B)$ preserves ${\cal H}^\ast$ is evident since 
$E(B){\cal H}^\ast=E(B\cap B^\ast){\cal H}\subset {\cal H}^\ast$.\\
$\Box$\\
Lemma \ref{la0.4} shows that any Hilbert space $\cal H$ is reducible 
with respect to a given p.v.m. $E$. The invariant subspaces 
${\cal H}^\ast$ are defined only with respect to $E$ and do not depend on 
the choice of a system of cyclic vectors $\Omega_n$.
\begin{Corollary} \label{col0.1} ~~~\\
Let $E'$ be a p.v.m. commuting with the p.v.m.
$E$, that is, $[E(B),E'(B')]=0$ for all $B,B'\in {\cal B}$. Then 
the spaces ${\cal H}^\ast$ defined with respect to $E$ are preserved by 
$E'$.
\end{Corollary}
This follows trivially from $E'(B){\cal H}^\ast=E'(B)E(B^\ast){\cal H}
=E(B^\ast)E'(B){\cal H}\subset {\cal H}^\ast$.\\
\\
The total 
decomposition $\mu=\mu^{pp}+\mu^{ac}+\mu^{cs}$ gives rise to a 
corresponding breakup of the spectrum $\sigma(a)$. 
\begin{Definition} \label{def0.5} ~\\
One defines the pure point spectrum $\sigma^{pp}(a)$ as 
the set of eigenvalues 
of $a$. This set may not be closed, however, 
$\sigma^\ast(a):=\sigma(a_{|{\cal H}^\ast})$ is closed. $\sigma^c(a):=
\sigma^{ac}(a)\cup \sigma^{cs}(a)$ is called the continuous spectrum. 
\end{Definition}
The three sets may not be disjoint and only $\overline{\sigma^{pp}(a)}\cup
\sigma^{ac}(a)\cup\sigma^{cs}(a)=\sigma(a)$. Roughly speaking, 
${\cal H}^{pp},\;{\cal H}^{ac},\;{\cal H}^{cs}$ correspond to bound,
scattering and states without physical interpretation respectively and a 
good deal of work in the spectral analysis of a given self -- adjoint 
operator is focussed on proving that $\sigma^{cs}(a)=\emptyset$.
A sufficient criterion is that there is a dense set $D$ of vectors 
$\psi$ such that for each $x\in \Rl$ the function $z\mapsto 
<\psi,R(z)\psi>$
is bounded as $z\to x$ where the bound is uniform, for given 
$\psi$, as $x$ takes values in any open interval. Here $R(z)=(a-z)^{-1}$
is the resolvent of $a$. The decomposition of the spectrum made above 
should not be confused with the disjoint decomposition into the discrete 
spectrum $\sigma^d(a)$ and the essential spectrum $\sigma^e(a)$ which are 
defined as the subset of $\sigma(a)$ consisting of the points $x$ such 
that $E((x-\epsilon,x+\epsilon))$ is a projection onto a finite or 
infinite subspace of ${\cal H}$ for any $\epsilon>0$. It is not difficult 
to show that 
$\sigma^d(a)$ consists of the isolated eigenvalues of finite multiplicity 
and that $\sigma^e(a)$ contains $\sigma^c(a)$, the limit points of 
$\sigma^{pp}(a)$ and the eigenvalues of infinite multiplicity. In 
particular it contains embedded eigenvalues, i.e. those which are also
part of the continuous spectrum.

\subsection{Direct Integral Representations and Constraint Quantization}
\label{s0.5}

By the theory just reviewed in section \ref{s0.3}, 
given a self -- adjoint operator $a$ with p.v.m. $E$ on a 
Hilbert space ${\cal H}$ we find a unitarily equivalent representation 
on a direct integral Hilbert space ${\cal H}_{\mu,N}$ such that 
$V E(B)=F(B) V$ for all Borel sets $B\in {\cal B}$ in $X=\Rl$ and the 
type $[\mu]$
and the multiplicity function $N$ are uniquely determined $\mu-$a.e.
up to unitary equivalence ($V:\;{\cal H}\to {\cal H}_{\mu,N}$ is the 
corresponding unitary map). We have by the spectral theorem for 
$\psi\in D(a)$ 
\be \label{0.39} 
<\psi,V a V^{-1}\psi'>_{{\cal H}_{\mu,N}}
=\int_\Rl \; x \;d<\psi,F(x)\psi'>_{{\cal H}_{\mu,N}} 
=\int_\Rl \; x \;d\mu(x)\; <\psi(x),\psi'(x)>_{{\cal H}_x} 
\ee
hence $V a V^{-1}$ is represented as mutiplication by $x$ on ${\cal H}_x$.

The kernel of $a$ is therefore the Hilbert space ${\cal H}_{x=0}$ and 
if $a$ is a constraint operator, we identify ${\cal H}_{phys}:={\cal 
H}_{x=0}$ with the physical Hilbert space. However, this prescription 
is too naive for the following reasons: 
\begin{itemize}
\item
Suppose that $\mu=\mu^c$ has no pure points. Then the set $\{x=0\}$
has $\mu-$measure zero and there is no harm, as far as the unitary 
equivalence with the direct integral representation is concerned, in 
choosing $N(x=0)$ arbitrarily, in particular setting $N(x=0)=0$ is 
allowed. 
In other words, in the case of only continuous spectrum the prescription 
is ambiguous. 
\item
Suppose that $x=0$ is an embedded eigenvalue, in other words, 
$\mu=\mu^{pp}+\mu^c$ with $\mu^{pp}(\{0\})>0$ and 
$\mu^c((-\epsilon,\epsilon))>0$ for all $\epsilon>0$.  
By the Radon -- Nikodym theorem $\mu_n(B)=\int_B \rho_n d\mu$ for 
non -- negative $\rho_n\in L_1(\Rl,d\mu)$. Since we have defined $N(x)$ 
as the number of $n$ such that $\rho_n(x)>0$ we are interested in the 
value of $\rho_n(0)$. Since $x=0$ is an eigenvalue there is at least 
one $n$, say $n=n_0$ such that $\mu_{n_0}(\{0\})>0$. Consequently 
$\mu(\{0\})\ge c_{n_0} \mu_{n_0}(\{0\})>0$. It follows that 
$\mu_n(\{0\})=\rho_n(0) \mu(\{0\})$. Thus, if $\mu_n^{pp}=0$ then 
we conclude $\rho_n(0)=0$ even if
$\mu_n((-\epsilon,\epsilon))=\mu_n^c((-\epsilon,\epsilon))>0$ for all
$\epsilon>0$. Thus we see that the presence of a single zero eigenvector 
would delete all generalized zero eigenvectors as one can see by comparing 
the situation with the one that resulted from deleting from $\mu=\sum_n
c_n \mu_n$ all terms corresponding to those $n$ such that $0\in P_n$.  
\end{itemize}
To see that it is physically wrong to suppress the continuous spectrum in 
case of embedded zero eigenvalues, consider the operator 
$C=p_1\otimes p_2$ on $L_2(\Rl,dx_1)\otimes L_2(S^1,dx_2)$ where 
zero is both an eigenvalue (corresponding to the eigenfunction 
$e_m(x_2):=e^{im x_2},\;m\in \Zl$ of $p_2$ with $m=0$) and a generalized 
eigenvalue
(corresponding to the generalized eigenfunction $f_k(x_1)=e^{ikx_1},\;k\in 
\Rl$ 
of $p_1$ with k=0). Classically $C=0$ corresponds to a particle 
moving on a cylinder which is constrained to move parallel to the 
direction 
of its axis or perpendicular to it. Thus one expects that the physical 
Hilbert space is spanned (in the appropriate sense) by vectors of the 
form $f_0\otimes \psi_2+\psi_1\otimes e_0$ with $\psi_j\in {\cal H}_j$
arbitrary which is isomorphic to 
${\cal H}_1\oplus {\cal H}_2$. However, the naive prescription would lead 
to the following result: Switching to the momentum representation with 
respect to $p_1$ we choose as our cyclic system of vectors the 
$\Omega_{0,m}:=H_0\otimes e_m,\;m\in\Zl-\{0\}$ and  
$\Omega_{n,0}:=H_n\otimes e_0,\;n\in\Nl$ where $H_n$ is the orthonormal  
basis of ${\cal H}_1=L_2(\Rl,dk)$ consisting of Hermite 
functions. The spectral measures are computed as 
\ba \label{0.43}
\mu_{0,m}(B) &= & <\Omega_{0,m},\chi_B(C)\Omega_{0,m}>
=\int_\Rl dk \chi_B(m k) |H_0(k)|^2 =\mu_{0,-m}(B)   
\nonumber\\
\mu_{n,0}(B) &= & <\Omega_{n,0},\chi_B(C)\Omega_{n,0}>
=\int_\Rl dk \chi_B(0) |H_0(k)|^2=\chi_B(0)
\ea
We see that $\mu_{0,m}=\mu_{0,m}^{ac}$ and $\mu_{n,0}=\mu_{n,0}^{pp}$
The total measure can be chosen to be 
\be \label{0.44}
\mu(B)=\frac{1}{3}[\sum_{m=1}^\infty 2^{-m} [\mu_{0,m}(B)+\mu_{0,-m}(B)]
+\frac{1}{2} \sum_{n=0}^\infty 2^{-n} \mu_{0,n}(B)]
=\frac{1}{3}[2\sum_{m=1}^\infty 2^{-m} \mu_{0,m}(B)+\chi_B(0)]
\ee
The Radon -- Nikodym derivatives are 
\ba \label{0.45}
\mu_{0,m}(B) &=& \int_B \rho_{0,m}(x) d\mu(x)
=\frac{1}{3}[2\sum_{k=1}^\infty 2^{-k} 
\int_B \rho_{0,m}(x) d\mu_{0,k}(x)+\rho_{0,m}(0)\chi_B(0)]
\nonumber\\
\mu_{n,0}(B) &=& \int_B \rho_{n,0}(x) d\mu(x)
=\frac{1}{3}[2\sum_{k=1}^\infty 2^{-k} 
\int_B \rho_{n,0}(x) d\mu_{0,k}(x)+\rho_{n,0}(0)\chi_B(0)]
\ea
Now choosing $B=(-\epsilon,\epsilon)$ and letting $\epsilon\to 0$ we 
deduce $\rho_{0,m}(0)=0$ and $\rho_{n,0}(0)=3>0$. Thus
it would follow from our naive prescription that ${\cal H}_{phys}
={\cal H}_{x=0}$ is spanned by the $\Omega_{n,0}$ and thus we would miss 
out completely the contribution from the continuous spectrum: Quantum 
mechanically the particle 
would only be allowed to move along the axis of the cylinder while 
classically it may also wrap around the cylinder. This is clearly 
physically wrong.

Thus, the naive prescription is ambiguous in the case that zero is only in 
the continuous 
spectrum, wrong in the case that zero is an embedded eigenvalue  
and unambiguous only if zero is an isolated eigenvalue in which case 
however the whole machinery of the direct integral is not needed at all
because then ${\cal H}_{x=0}\subset {\cal H}$ and the physical inner
product coincides with the kinematical one.

To improve this we prescribe the following procedure:\\
\\
{\bf PRESCRIPTION:}\\ 
The obvious solution to the second problem is to use the 
orthogonal decomposition of ${\cal H}$ into the pieces 
${\cal H}^{pp},\;{\cal H}^{ac},\;{\cal H}^{cs}$ derived in section 
\ref{s0.4} before applying the direct 
integral decomposition. As we have shown, this preliminary decomposition 
only depends on the type $[E]$ of $E$ and reduces both the p.v.m.
$E$ of the self -- adjoint operator $a$ of interest as well 
as the p.v.m.'s $E'$ of self -- adjoint operators $b$ which 
commute with $a$. (Notice that self -- adjoint operators $a,b$ are said to 
commute if and only if all their spectral projections commute. This avoids 
domain questions of unbounded operators.) We may therefore apply all the 
results of the previous section to the pieces individually. 

In practice, we then have to compute the individual 
Radon -- Nikodym derivatives 
\be \label{0.46}
\rho_n^\ast:=d\mu_n^\ast/d\mu^\ast,\;\;\mu^\ast=\sum_n c_n^\ast \mu_n^\ast  
\ee
While $\rho_n^{pp}(x)$ is unambiguously defined, 
$\rho_n^{ac}(x),\rho_n^{cs}\ge 0$ are 
only defined $\mu^{ac},\mu^{cs}-$a.e. respectively. In order to fix this 
ambiguity we need additional physical input. Namely, first of all we add 
the the requirement that a complete subalgebra\footnote{The corresponding 
classical 
functions should separate the points of the reduced phase space and form 
a closed Poisson subalgebra.} 
of bounded Dirac (weak or strong) observables be 
represented irreducibly as self adjoint operators on the physical Hilbert 
space\footnote{We may allow quantum corrections to the classical Poisson
algebra of these observables.}.

Let us see what this implies given the structure already 
available: By theorem \ref{th0.5} ii), different choices of 
$\Omega^\ast_n$ and 
$c^\ast_n$ in $\mu^\ast=\sum_n c_n^\ast \mu^\ast_n$ lead to unitarily 
equivalent direct integral representations ${\cal H}_{\mu^\ast,N^\ast}$
where the type $[\mu^\ast]$ and the $\mu^\ast-$class of the function 
$N^\ast$ are 
unique. By theorem 
\ref{th0.2} i) there exists a unitary operator mediating between these 
two realizations of the form  
$V^\ast:\;{\cal H}_{\mu^\ast,N^\ast} \to {\cal 
H}_{\mu^{\ast\prime},N^{\ast\prime}};\;(V\psi)(x)=\sigma_\ast(x)
U(x)\psi(x)$ where $U(x):\;{\cal H}^{\oplus\ast}_x\to {\cal 
H}^{\oplus\ast\prime}_x$ 
is a measurable, fibre preserving unitarity and 
$\sigma_\ast^2(x)=[d\mu^\ast/d\mu^{\ast\prime}](x)$. Moreover, given a 
bounded, self -- adjoint strong Dirac
observable, $A$ we see from theorem \ref{th0.3} ii) that it gives rise to 
a measurable, bounded, self -- adjoint and fibre preserving operator of 
the form $(a\psi)(x)=a(x) \psi(x)$ in any direct integral representation.
It follows that $(Va V^{-1})(x)=U(x) a(x) U^{-1}(x)$, in other words, 
different direct integral representations give rise to unitarily 
equivalent representations of strong Dirac observables in $\mu-$a.e.
fibre. It follows that once we have 
chosen the $\rho^\ast_n(0)$ for one choice of $\Omega^\ast_n,\; c^\ast_n$
we may fix them for all others by requiring that the representations of 
the strong Dirac observables be exactly unitarily equivalent in the fibre 
$x=0$. This means that 
in particular the number $N^\ast(0)$ of those $n$ for which 
$\rho^\ast_n(0)>0$ is fixed for all those choices of $\Omega^\ast_n, 
c^\ast_n$. 
Notice that the positive constants $K_\ast:=\sigma_\ast(0)$ by 
which the norms of $||\psi(0)||_{{\cal H}^{\oplus\ast}_0},\; 
\sigma_\ast(0)||U(0)\psi(0)||_{{\cal H}^{\oplus\ast\prime}_0}$
of the induced physical inner products differ are irrelevant for 
strong Dirac observables because 
they drop out in the normalization of states. However, they play a role 
for weak Dirac observables which may mix the sectors ${\cal H}^\ast$.

Thus we may
restrict attention to one choice $\Omega^\ast_n,\;c^\ast_n$ and are left 
with the choice of the representatives $\rho^\ast_n(0)$ for those 
fixed data (and the three undetermined numbers 
$K_\ast,\;\ast=\{pp,ac,cs\}$ which we should have added anyway because 
it is anway ad hoc to equip the physical Hilbert space with the inner 
product of ${\cal H}^{\oplus\ast}_0$ rather than any positive scalar
multiple thereof).  
Fortunately, as we will see in the next section, either choice of 
$\rho^\ast_n(0)$
induces a self -- adjoint representation of bounded strong Dirac 
observables. 
Moreover, since weak Dirac observables can be characterized 
by the fact that they preserve at least the fibre $x=0$, they are 
therefore equivalent to strong Dirac observables as far as the fibre 
$x=0$ is concerned and hence the self -- adjointness criterion 
also does not fix the remaining 
ambiguity, although they may fix the constants $K_\ast$. However, non -- 
trivial restrictions arise from the fact that we want a reresentation of 
the {\it algebra} of observables. This will in general prohibit to alter 
the $\rho^\ast_n(0)$ arbitrarily.

If this still does not fix the ambiguity,
we must look for other criteria, such as whether the resulting 
physical Hilbert space admits a sufficient number of semiclassical 
states\footnote{In some sense this granted by Fell's theorem
\cite{1.6} once we have a representation on the physical Hilbert space of 
the $C^\ast$ algebra 
of our preferred algebra of bounded Dirac observables.}.
Fortunately, as suggested 
by the examples, the a priori knowledge of good physical semiclassical 
states or the algebra of Dirac observables seems not 
to be necessary but rather one can take the following practical approach:
We choose a minimal set of $\Omega_n$ (which is always 
possible and measure theoretically unique, see below) and for such a 
minimal set we choose an everywhere non -- negative representative,
from the equivalence 
class of the measurable functions which equal $\rho_n^{ac}$ 
$\mu^{ac}-$a.e. (and similar for $\rho_n^{cs}$), which 
is continuous at $\lambda=0$ from the right, if such a representative 
exists. If it does, then $\rho_n^c(0):=\lim_{x\to 0+} \rho_n^c(x)$ is well 
defined. If such a representative does not exist, a case which was not 
encountered so far in the examples we studied, then we must really resort 
to the physical criteria or, if even that does not fix the ambiguity, 
we must adopt an ad hoc prescription such as 
arbitrarily setting $\rho^c_n(0)=0$ in this case. Further restrictions may 
come from the irreducibility criterion\footnote{The set of bounded 
operators on a Hilbert space is always represented irreducibly. However, 
here the question is whether the subset of bounded operators induced 
from the kinematical Hilbert space is represented irreducibly.}.

We stress that the fundamental, physical prescription is always the 
irreducible self 
-- adjoint representation of a complete subalgebra of Dirac observables 
and a good 
semiclassical behaviour.  Ideally one would want to show that the 
freedom in the more practical continuity prescription 
is equivalent to the freedom left in the physical 
prescription. In the examples encountered that happened to be the case 
but in general there seems to be little known about the relation between
these two prescriptions. It is conceivable that in general 
the physical representations induced from a kinematical one are 
simply not uniquely determined. We will come back to this issue 
in section \ref{0.7} where we compare with the amount of ambiguities in 
other approaches.\\ 
This ends our prescription.\\ 
\\
As follows from the proof of the spectral theorem which uses the Riesz 
Markov theorem, all the measures $\mu_n^\ast,\mu^\ast$ are regular, finite 
Borel measures on $\Rl$  
and therefore we may apply Lusin's theorem \cite{Rudin} which says 
that $\rho_n^{ac},\;\rho_n^{cs}$ can be approximated, up to sets of 
arbitrarily 
small $\mu^{ac},\mu^{cs}$ measure, by a continuous function. Hence the 
continuity part in our  
prescription makes sense. In practice the 
continuous singular part is mostly absent and then it will be sufficient 
to choose a representative, defined a.e. with respect to Lebesgue measure, 
which is maximally continuous and non -- negative. Also in practice the 
representatives that one computes are naturally non -- negative 
everywhere. 

The reason for why we choose the number of $\Omega_n$ to be 
minimal is in order to remove the following, trivial ambiguity:
Suppose for instance that $\cal H$ even has a cyclic vector $\Omega_0$. 
Let 
$I_n,\;n=1,..,m$ be a system of mutually disjoint intervals whose union is 
$\Rl$ and set $\Omega_n:=E(I_n)\Omega_0/||E(I_n)\Omega_0)||$. Then the 
$\Omega_n$ provide an orthonormal cyclic system as well whose total 
measure is equivalent to the spectral measure of $\Omega_0$. We have 
$\mu_n(B)=<\Omega_0,E(B\cap I_n)\Omega_0>$ hence $\rho_n=\chi_{I_n}$ 
for $n=1,..,m$ while $\rho_0=1$. We see that we can make the Radon -- 
Nikodym derivatives arbitrarily discontinuous at any values by an 
unfortunate, that is, redundant choice 
of $\Omega_n$ and to avoid that it is obviously necessary to minimize the 
number of necessary $\Omega_n$. This number is given by the maximal 
multiplicity $M=\mu-\max(N)$ of the function $N$. That this is always 
possible is the content of the subsequent lemma:
\begin{Lemma} \label{la0.4a} ~\\
The cyclic system $\Omega_n$ can be chosen in such a way that with
$\Omega=\sum_{n=1}^M 2^{-n/2} \Omega_n/\sqrt{\sum_{n=1}^M 2^{-n}}$ we have
$[\mu_\Omega]=[\mu_{\Omega_1}]\ge [\mu_{\Omega_2}]\ge ...$ where the 
notation $[\mu]\ge [\nu]$ means that $\nu'$ is absolutely continuous 
with respect to $\mu'$ for any $\mu'\in [\mu],\;\nu'\in [\nu]$. 
Moreover, the above types are uniquely determined.
\end{Lemma}
Proof of lemma \ref{la0.4a}:\\
Let $\Omega_1$ be any vector such that its spectral measure has maximal  
type, see lemma \ref{la0.1}. Suppose now that we have found already
mutually orthogonal $\Omega_1,..,\Omega_n$ such that 
the ${\cal H}_k:=\overline{\mbox{span}\{E(B)\Omega_k;\;B\in {\cal B}\}}$
are mutually orthogonal and such that 
$[\mu_{\Omega_1}]\ge..\ge [\mu_{\Omega_n}]$. Put 
${\cal H}^{(n)}:=\oplus_{k=1}^n {\cal H}_k$. We have 
${\cal H}^{(k)}\subset {\cal H}^{(k+1)}$ for $k=1,..,n-1$ and  
${\cal H}^{(k+1)\perp} \subset {\cal H}^{(k)\perp}$ for $k=0,..,n-1$ 
where we have set ${\cal H}^{(0)\perp}:={\cal H}$. Let $[E^\perp_k]$ be 
the 
maximal type of the spectral measures 
$\mu_\psi,\;\psi\in {\cal H}^{(k)\perp}$, that is $[\mu_\psi]\le 
E^\perp_k$ for all $\psi\in {\cal H}^{(k)\perp}$. Since 
${\cal H}^{(k+1)\perp} \subset {\cal H}^{(k)\perp}$ we also have 
$[\mu_\psi]\le [E^\perp_k]$ for all $\psi\in {\cal H}^{(k+1)\perp}$. It 
follows that $[E^\perp_{k+1}]\le [E^\perp_k]$ for $k=0,..,n-1$ where of 
course $[E^\perp_0]=[E]$. We now make the additional induction assumption 
that $[\mu_{\Omega_k}]=[E^\perp_{k-1}]$ for $k=1,..,n$ which is obviously 
satisfied for 
$k=1$. Choose some $\Omega_{n+1}\in {\cal H}^{(n)\perp}$ of maximal 
type, i.e. $[\mu_{\Omega_{n+1}}]=[E^\perp_n]$. Then obviously
$[E^\perp_n]=[\mu_{\Omega_{n+1}}]\le [\mu_{\Omega_n}]=[E^\perp_{n-1}]$
as claimed. 

To see that the measure classes $[\mu_{\Omega_n}]$ are uniquely 
determined, consider the supports
$S_n:=\{x\in X;\; \rho_n(x):=d\mu_{\Omega_n}(x)/d\mu_{\Omega_1}(x)>0\}$. 
It follows 
that up to $\mu_{\Omega_1}-$measure zero sets we have $S_{n+1}\subset 
S_n$.
Then $X_n:=S_n-S_{n+1}$ coincides with the set 
$N^{-1}(n)=\{x\in X; \rho_k(x)>0 \mbox{ for precisely }n \mbox{ of the 
}k\}$. Since $N$ is uniquely determined up to sets of $\mu-$measure zero,
so are the $X_n$ and thus the $S_n:=\cup_{k=1}^n X_k$. Thus 
$[\mu_{\Omega_n}]$ is uniquely determined.\\
$\Box$\\
\\
Let us see how our prescription affects the direct 
integral and dimension function $N$. We write for $\Psi\in {\cal H}$
\ba \label{0.47}
||\Psi||^2 &=& 
\sum_{n=1}^M \int_X\; d\mu_n\; |\Psi_n(x)|^2
=\sum_{\ast=pp,ac,cs} \sum_{n=1}^M \int_X\; d\mu_n^\ast\; 
|\Psi_n(x)|^2
\nonumber\\
&=& 
\sum_{\ast=pp,ac,cs}
\sum_{n=1}^M \int_X\; d\mu^\ast\; \rho_n^\ast(x)|\Psi_n(x)|^2
\nonumber\\
&=& 
\sum_{\ast=pp,ac,cs}
\int_X\; d\mu^\ast\; [\sum_{n=1}^M
|\sqrt{\rho_n^\ast(x)}\Psi_n(x)|^2]
\ea
In other words, all of section \ref{s0.3} applies, the only difference 
being that now we have a unitary map between ${\cal H}^\ast$ and 
${\cal H}_{\mu^\ast,N^\ast}$ where 
$N^\ast(x)$ are defined $\mu^\ast-$a.e.  
as the number of $n$ such that $\rho_n^\ast(x)>0$.
If we compare (\ref{0.47}) with the unitarity equivalence between 
${\cal H}$ and ${\cal H}_{\mu,N}$ given as in section \ref{s0.3} by
\be \label{0.48}
||\Psi||^2=\int_X\; d\mu\; [\sum_{n=1}^M |\sqrt{\rho_n(x)}\Psi_n(x)|^2]
=\sum_{\ast=pp,ac,cs} \int_X\; d\mu^\ast\; [\sum_{n=1}^M
|\sqrt{\rho_n(x)}\Psi_n(x)|^2]
\ee
then we conclude that $\rho_n^\ast=\rho_n$ $\mu^\ast-$a.e. 
Thus $\rho_n^{pp}$ can differ from $\rho_n$
everywhere except at the pure points $p\in P$ of $\mu$ while 
$\rho_n^{ac},\;\rho_n^{cs}$ 
may differ from $\rho_n$ in particular at the pure points of $\mu$. It is 
precisely this fact which allows us to repair the second problem 
mentioned above. Without loss of generality and in order to be specific we 
may choose $\rho_n^{pp}(x)=0$ for $x\not\in P$ while 
$\rho_n^{pp}(x)=\rho_n(x)$ is fixed for $x\in P$. Thus 
$\rho_n^{pp}(x)=\sum_{p\in P} \delta_{x,p} \rho_n(p)$. 
For $\rho_n^{ac},\;\rho_n^{cs}$ 
we use our prescription spelled out above. Notice that it is possible and 
of practical advantage to split the set pf $\Omega_n$ into 
the respective sets of $\Omega_n^\ast\in {\cal H}^\ast,\;n=1,2,..,M^\ast$
and to define $\mu^\ast=\sum_n c_n^\ast \mu^\ast_n$ and 
$\Psi=\sum_{n,\ast} \Psi^\ast_n(a) \Omega_n^\ast$. We will assume to have
done that in what follows. 

The physical Hilbert space is then evidently the direct sum
${\cal H}_{phys}:={\cal H}^{pp}_{x=0}\oplus 
{\cal H}^{ac}_{x=0}\oplus {\cal H}^{cs}_{x=0}$.
Notice that the physical Hilbert space can be represented as an $\ell_2$
space consisting of sequences of complex numbers 
$z^\ast_n,\;n=1,..,M^\ast$ subject to $\sum_{n,\ast} K_\ast |z_n^\ast|^2 
\rho^\ast_n(0)<\infty$. It is clear that different choices of 
$\Omega^\ast_n,\;c^\ast_n$ 
result in unitarily equivalent Hilbert spaces
since by our prescription the numbers $N^\ast(0)$ of those $n$ with 
$\rho_n^\ast(0)>0$ is fixed.

Whether the superselection structure concerning the spectral types with 
respect to the strong Dirac observables remains intact if we also 
allow weak Dirac observables cannot be answered in general and will 
probably depend on the concrete constraint operator under investigation.
See e.g. \cite{MMT} where this actually happens in a different context.\\
\\ 
Let us verify that our prescription leads to the correct answer in the 
example discussed above: We have 
\be \label{0.49}
\mu_{0,m}(B)=\int dk \chi_B(mk) |H_0(k)|^2
=\frac{1}{|m|}\int_B dx  |H_0(x/m)|^2
\ee
hence $d\mu_{0,m}(x)/dx=|H_0(x/m)|^2/|m|$ and 
$d\mu^c(x)/dx=2/3\sum_{k=0}^\infty 2^{-k} d\mu_{0,k}(x)/dx$ so that 
\be \label{0.50}
\rho_{m,0}^c(x)=\frac{3|H_0(x/m)|^2/|m|}{2 
\sum_{k=0}^\infty 2^{-k}/|k| |H_0(x/k)|^2}
\ee
This function is already continuous, even smooth and actually everywhere
positive, in particular
\be \label{0.50a}
\rho_{m,0}^c(0)=\frac{3}{2|m|\sum_{k=0}^\infty 2^{-k}/k}>0
\ee
for all $m\not=0$. Thus $N^c(x)=N^{pp}(x)=|\aleph|$ have countable 
cardinality independent of $x$.

Introducing orthonormal bases $e_{m,0};\;m\in \Zl-\{0\}$ 
in the associated $l_2$ space and likewise $e_{0,n};\;n=0,1,2,..$ 
we have that under the unitary map 
$V:\;{\cal H}\to {\cal H}_{\mu^{pp},N^{pp}}\oplus {\cal H}_{\mu^c,N^c}$
\be \label{0.51}
(V\Psi)(0)=\sum_{m\not=0} \Psi_{m,0}(0) \sqrt{\rho_{m,0}(0)} e_{m,0}
\oplus
\sum_n \Psi_{0,n}(0) \sqrt{\rho_{0,n}(0)} e_{0,n}
\ee
In particular for 
$\Psi=\Omega_{m,0},\;\Psi_{k,0}=\delta_{k,m},\;\Psi_{0,k}=0$ and 
$\Psi=\Omega_{0,n},\;\Psi_{0,k}=\delta_{k,n},\;\Psi_{k,0}=0$ 
we find 
\be \label{0.52}
(V\Omega_{m,0})(0)=\sqrt{\rho_{m,0}(0)} e_{m,0} \mbox{ and } 
(V\Omega_{0,n})(0)=\sqrt{\rho_{0,n}(0)} e_{0,n} 
\ee
which shows that the heuristic expectation is correct, namely that the 
span of the $\Omega_{m,0}$ which is isomorphic to the orthogonal 
complement of the vector $1$ in the Hilbert space 
$L_2(S^1,dx_2)$ is isometric isomorphic to the 
span of the $e_{m,0}$ while the span of the $\Omega_{0,n}$ which is 
isomorphic to the Hilbert space 
$L_2(\Rl,dx_1)$ is isometric isomorphic to the 
span of the $e_{0,n}$. Moreover, these two physical Hilbert spaces are
realized as direct sums. Notice that we could attribute the vector 
$\Omega_{0,0}$ also to $L_2(S^1,dx_2)$ but then we would have to subtract 
it from $L_2(\Rl,dx_1)$. This effect is related to the fact that the point 
$p_1=p_2=0$ also classically plays a special role: Namely the reduced 
phase space with respect to the constraint $C=p_1 p_2$ is as follows:
The constraint surface is not a manifold but a variety of 
varying dimension consisting of the five disjoint pieces 
$S^\pm_1=\{(x_1,x_2,\pm p_1>0,p_2=0)\},\;S^\pm_2=\{(x_1,x_2,\pm 
p_2>0,p_1=0)\},\;S_0=\{(x_1,x_2,p_1=0,p_2=0)\}$. The constraint generates 
gauge motions on each of these pieces except for $S_0$ and leads to the 
reduced phase space consting of the disjoint pieces
$P^\pm_1=\{(x_1,\pm p_1>0)\},\;P^\pm_2=\{(x_2,\pm 
p_2>0)\},\;P_0=\{(x_1,x_2)\}$. Notice that $P_0$ has a degenerate symplectic 
structure and thus should be discarded. But even then we see that the 
reduced phase space is not the union of two cotangent bundles over 
$\Rl$ but rather of four cotangent bundles over $\Rl_+$. This non -- 
trivial topology is reflected in the above direct sum which is not the 
direct sum of the two Hilbert spaces corresponding to the union of two 
topologically trivial contangent bundles. We will not dwell further on 
this point, the discussion is just to reveal that the unusual form of 
${\cal H}_{phys}$ is not surprising.

\subsection{Explicit Action of Dirac Observables on the 
Physical Hilbert Space}
\label{s0.6}

As we have explained we may focus attention on either of the sectors
${\cal H}^\ast$ seperately. We will drop the $\ast$ for the purposes of 
this section.\\
\\
Let $E$ be the p.v.m. of a self -- adjoint constraint 
operator $a$ and let $E'$ be the p.v.m. of a strong 
Dirac observable $b$. Let $V:\;{\cal H}\to {\cal H}_{\mu,N}$ be an 
associated direct integral representation based on a cyclic system 
of vectors $\Omega_n$. Let $f$ be a measurable function and 
$\Psi'\in D(f(b))$. From section \ref{s0.3} we know that $V f(b) V^{-1}$ 
is fibre preserving and determines $\mu-$a.e. uniquely an operator 
$[f(b)](x)$ on ${\cal H}_x$. This applies in particular to the spectral
projections $E'(\lambda):=E'((-\infty,\lambda])$. By the spectral theorem 
\ba \label{0.53}
<\psi,V f(b) V^{-1} \psi'>_{{\cal H}_{\mu,N}}
&=& \int_\Rl \; f(\lambda)\; 
d<\psi,V E'(\lambda) V^{-1} \psi'>_{{\cal H}_{\mu,N}}
\nonumber\\
&=& \int_\Rl \; f(\lambda)\; \int_\Rl \; d\mu(x)\;
d<\psi(x),[E'(\lambda)](x)\psi'>_{{\cal H}_x}
\nonumber\\
&=& \int_\Rl \; d\mu(x)\; \int_\Rl \; f(\lambda)\; 
d<\psi(x),[E'(\lambda)](x)\psi'>_{{\cal H}_x}
\ea
whence $\mu-$a.e.
\be \label{0.54}
[f(b)](x)=\int_\Rl \; f(\lambda)\; d_\lambda [E'(\lambda)](x)
\ee
Thus we only need to know $[E'(\lambda)](x)$. There are measurable 
functions $x\mapsto G^\lambda_{nm}(x)$ such that 
\be \label{0.55}
E'(\lambda)\Omega_m=\sum_n\; G^\lambda_{mn}(a)\;\Omega_n
\ee
Therefore for all $\Psi,\Psi'\in {\cal H}$
\ba \label{0.56}
<\Psi,E'(\lambda)\Psi'>_{{\cal H}} 
&=&
=\sum_n \int\; d\mu_n(x)\; \overline{\Psi_n(x)} \;\sum_m \;
G^\lambda_{mn}(x) \Psi'_m(x)
\nonumber\\
&=&
\int\; d\mu(x)\; \sum_{n\in M(x)} \rho_n(x) \overline{\Psi_n(x)} \;\sum_m 
\;
G^\lambda_{mn}(x) \Psi'_m(x)
\ea
On the other hand 
\be \label{0.57}
(V\Psi)(x)=\psi(x)=\sum_{n\in M(x)} \sqrt{\rho_n(x)} \Psi_n(x) e_n(x)
\ee
with $e_n(x),\;n\in M(x)$ an orthonormal basis of ${\cal H}_x$ and 
$M(x)=\{n:\;\rho_n(x)>0\}$ as described in section \ref{s0.3}. Thus 
defining 
\be \label{0.58}
[E'(\lambda)](x) e_m(x)=:\sum_{n\in M(x)} ([E'(\lambda)](x))_{mn} e_n(x)
\ee
we have
\ba \label{0.59}
<\Psi,E'(\lambda)\Psi'>_{{\cal H}} 
&=&
\int\; d\mu(x)\; <\psi(x),[E'(\lambda)](x)\psi'(x)>_{{\cal H}_x}
\nonumber\\
&=&
\int\; d\mu(x)\; \sum_{m,n\in 
M(x)} \overline{\Psi_n(x)} ([E'(\lambda)](x))_{mn} \Psi'_m(x)
\sqrt{\rho_n(x)\rho_m(x)}
\ea
We conclude that $\mu-$a.e. 
\be \label{0.60}
([E'(\lambda)](x))_{mn}=
\chi_{M(x)}(m)\;\chi_{M(x)}(n)\;\sqrt{\frac{\rho_n(x)}{\rho_m(x)}}\;
G^\lambda_{mn}(x) 
\ee
In order to fix (\ref{0.60}) one must choose a representative 
$G^\lambda_{mn}(x)$ such that (\ref{0.60}) is self -- adjoint which 
is always possible $\mu-$a.e. by the results of section \ref{s0.3}.
For the pure point sector these numbers are uniquely determined while for 
the continuous sectors we will use our 
prescription to fix the freedom in (\ref{0.60}) at $x=0$. For instance we 
may 
reduce the freedom by insisting that  
$\lambda \to [E'(\lambda)](0)$ has to be a system of spectral projections
on ${\cal H}_x$. See below for the general case.\\
\\
In practice one is directly interested in the Dirac observables
$b$ themselves and thus one will try to choose the 
$\Omega_n$ to be in their common domain and as $C^\infty-$vectors of 
the constraint operator $a$. One can then 
directly determine the measurable functions $G_{mn}(a)$ via 
$b\Omega_m=\sum_n \; G_{mn}(a) \Omega_n$. The resulting expression 
for $(b(x))_{mn}$ then is 
analogous to (\ref{0.60}). Notice that for bounded,
strong, self adjoint Dirac observables the induced operator on the 
physical Hilbert space is bounded and self -- adjoint no matter how the 
$\rho_n(0)$ were chosen because the $(b(x))_{mn}$ are non vanishing only
if both $m,n\in M(x)$. This follows because by self -- adjointness 
of $b$ we have $\mu-$a.e. 
$\rho_m\overline{G_{nm}}=\rho_n G_{mn}$
and so we may choose a Hermitean representative $b_{mn}$.
Moreover, since all 
possible direct integral representations of a given Hilbert space 
induced by different choices of $\Omega_n, c_n$ are unitarily 
equivalent as we showed in section \ref{s0.5} inducing a measurable 
fibre preserving unitarity, we may always arrange that different such 
choices lead to unitarily equivalent induced representations on the 
physical Hilbert space. Interestingly, if we find strong, unitary 
Dirac Observables $u$ then we may simplify the spectral analysis because 
then the two vectors $\Omega_1$ and $\Omega_2:=U\Omega_1$ have the 
same spectral measures.\\
\\
The discussion of weak Dirac observables is more complicated because 
they are not necessarily fibre preserving. We will only sketch some ideas and 
reserve a complete discussion for future publications.
By definition, if we write 
$\Psi=\sum_{\ast,n} \Psi^\ast_n(\MCO) \Omega^\ast_n$ for measurable 
functions $\Psi^\ast_n$ then the direct integral representation of 
$\Psi$ is given by $(V\Psi)(x)=\psi(x)=\sum_{\ast,n} \sqrt{\rho^\ast_n(x)} 
\Psi^\ast_n(x) e^\ast_n$ where
$V:\;{\cal H}\to {\cal H}_{\mu,N}$ is the unitary 
operator which realizes $\cal H$ as a direct integral.
Weak bounded s.a. Dirac
observables are of the form 
$(V \hat{D} \Psi)(x)=\int d\nu^D(x') d(x',x)\psi(x')$ for some measure 
$\nu^D$ 
and some 
kernel $d(x',x):\;{\cal H}^\oplus_{x'}\mapsto {\cal H}^\oplus_x$. The 
classical condition 
for a weak Dirac observable $\{D,\{D,\MC\}\}_{\MC=0}=0$ translates into the 
condition that 
$A:=V [\hat{D},[\hat{D},\MCO]] V^{-1}$ should annihilate the fibre 
${\cal H}^\oplus_0$. In other words, if $a(x',x):\; 
{\cal H}^\oplus_{x'}\mapsto {\cal H}^\oplus_x$ is the kernel of $A$,
that is, $(A\psi)(x)=\int d\nu^A(x') a(x',x) \psi(x')$ then 
$a(0,x)=0$ for $\mu-$a.a. $x$. In terms of the measure 
$\nu^D$ and the kernel $d$ we have explicitly $\nu^A=\nu^D$ and 
$a(x',x)=\int d\nu^D(x^{\prime\prime}) 
d(x^{\prime\prime},x) d(x',x^{\prime\prime})[x+x^{\prime\prime}-2x']$. 
This condition is implied by $d(0,x)=0$ for $\nu^D-$a.a. $x$ which would 
mean that the fibre ${\cal H}^\oplus_0$ is preserved but not 
necessarily the individual sectors ${\cal H}^{\oplus\ast}_0$. This is 
in fact the only sensible choice if ${\cal H}^\oplus_0$ is to carry an 
induced representation of the weak Dirac observables. We conclude 
that $(V\hat{D}\Psi)(0)=\nu^D(\{0\}) d(0,0)(V\psi)(0)$ where 
$\nu^D(\{0\})\not=0$. 
%

In future publications we will elaborate more on representations of weak
Dirac observables and
investigate the question under which circumstances the superselection 
structure with respect to strong Dirac observables is destroyed by the 
weak ones. Notice, however, that also self -- adjointness of weak, 
bounded self -- adjoint Dirac observables 
cannot fix the ambiguity in the choice of the $\rho_n(0)$ since they must 
preserve the fibre $x=0$ as we just showed and are then automatically self 
-- adjoint there
for the same reason as the strong Dirac observables.\\
\\
We now will exhibit that the requirement of a self -- 
adjoint representation of the {\it algebra} of strong Dirac observables 
will impose severe constraints on the sets $M(x)=\{n\in 
\Nl;\rho_n(x)>0\}$. Let $D^j,\;j=1,2,3$ be strong Dirac observables on 
$\cal H$ with $D^1 D^2=D^3$
and $D^j\Omega_m=\sum_n D^j_{mn}(a) \Omega_n$. 
It follows for the measurable functions that $\sum_k D^2_{mk} D^1_{kn}=
D^3_{mn}$. The corresponding 
operators in the fibres are then given by 
$d^j(x)e_m=\sum_n d^j_{mn}(x) e_n$ where 
$d^j_{mn}(x)=\sqrt{\rho_n/\rho_m}(x)
\chi_{M(x)}(m)\chi_{M(x)}(n) D^j_{mn}(x)$. Now a short calculation reveals
\be \label{0.80}  
d^1(x) d^2(x)=\sqrt{\frac{\rho_n(x)}{\rho_m(x)}}
\chi_{M(x)}(m)\chi_{M(x)}(n) \sum_{k\in M(x)} D^2_{mk}(x) D^1_{kn}(x)
\ee
which coincides $\mu-$a.e. with
\be \label{0.81}  
d^3(x)=\sqrt{\frac{\rho_n(x)}{\rho_m(x)}}
\chi_{M(x)}(m)\chi_{M(x)}(n) D^3_{mn}(x)
=\sqrt{\frac{\rho_n(x)}{\rho_m(x)}}
\chi_{M(x)}(m)\chi_{M(x)}(n) \sum_k D^2_{mk}(x) 
D^1_{kn}(x)
\ee
The point is now that (\ref{0.80}) and (\ref{0.81}) differ by the fact 
that in (\ref{0.80}) the sum over $k$ is restricted to the set $M(x)$
while in (\ref{0.81}) it is not. Requiring that these two expressions 
coincide, at least at $x=0$, numerically rather than a.e. should impose 
restrictions on the choice of the representatives of $\rho_n(0), 
D^j_{mn}(0)$ and all other Dirac observables. Intuitively, this 
requirement will amount to choosing representatives $\rho_n(0)$ which are 
positive for a maximal number of $n$ so that we are not missing 
necessary terms while irreducibility will require to have a maximal
number of the $\rho_n(0)$ vanishing so that at least heuristically these 
two requirements have the tendency to restrict the freedom.

\subsection{Comparison Between Refined Algebraic Quantization (RAQ)
and the Direct Integral Decomposition (DID)}
\label{s0.7}

The main purpose of the Master Constraint Programme is to deal with 
situations where RAQ \cite{Marolf} fails: Namely in the case of an 
infinite dimensional set of constraints with no or little control on
the resulting group they generate or, even worse, when there is no group 
at all (the constraints close with non -- trivial structure functions on 
phase space rather than structure constants). In such situations there are 
presently only formal BRST procedures available \cite{X} which apply at 
most 
in the case of a finite number of degrees if freedom and which have not 
yet been shown to produce a non negative physical inner product.
On the other hand, the Master 
Constraint Programme offers a rigorous alternative {\it whose mathematics 
always works}. 

In the present section we would like to compare the direct integral 
decomposition (DID) method with the RAQ programme in a situation to which 
RAQ applies: This is the case of a {\it single} constraint or the  
Master Constraint considered as a {\it single} constraint.   

In its most general form, RAQ consists of the following steps 
\cite{Marolf}:
\begin{itemize}
\item[1.]
{\it Choose} a dense and invariant domain $\Phi$ 
for $a$. $\Phi^\ast$ is defined as the algebraic dual of $\Phi$, i.e. the 
set of all linear functionals on $\Phi$ equipped with the weak 
$^\ast$ topology of pointwise convergence.
\item[2.]
An element $F\in \Phi^\ast$ is said to be a physical ``state'' provided 
that $F[a^\dagger f]=0$ for all $f\in \Phi$. Denote the vector space of 
these 
generalized solutions by $\Phi^\ast_{phys}$.
\item[3.]
Turn (a subspace of) $\Phi^\ast_{phys}$ into a pre -- Hilbert space by 
supplementing it with an anti -- linear ``Rigging Map'' 
\be \label{0.61}
\eta:\;\Phi \to \Phi^\ast_{phys};\;f\mapsto \eta(f);\;\;
<\eta(f),\eta(f')>_{phys}:=(\eta(f'))[f]
\ee
and then complete it with respect to the sesqui -- linear form 
$<.,.>_{phys}$ to obtain a physical Hilbert space ${\cal H}_{phys}$ (possibly 
after dividing out a null space).
In order that $\eta$ be a rigging map, (A) (\ref{0.61}) must be a positive 
semi -- definite 
sesqui -- linear form and, moreover, (B) for any strong Dirac observable 
defined on $\Phi$ we should have $b'\eta(f)=\eta(bf)$ where $b'$ is the 
dual of $b$ defined on $\Phi^\ast$ via $(b' F)[f]:=F(b^\dagger f)$ and 
$b^\dagger$ is the adjoint of $b$ on ${\cal H}$. One says that $b$ commutes 
with the rigging map. It is easy to see that condition (B) implies that 
symmetric operators on ${\cal H}$ are promoted to symmetric operators on 
${\cal H}_{phys}$. 
\end{itemize}
A heuristic procedure for constructing $\eta$ from $a$ is to set 
$(\eta(f))[f']:=<f,\delta(a) f'>$ where the $\delta-$distribution 
is formally defined via the spectral theorem as 
$\delta(a)=\int\;\delta(\lambda) dE(\lambda)$. It is clear that this 
formally solves the requirements on $\eta$ to qualify as a rigging map,
however, the meaning of the $\delta-$distribution must be made more 
precise and depends on the spectral properties of $a$. In fact, the direct 
integral representation of $\cal H$ now enables us to precisely do 
that as follows:\\
Recall the decomposition ${\cal H}=\sum_{\ast=pp,ac,cs} {\cal H}^\ast$.
The ``operator'' $\delta(a)$ is reduced by this decomposition and we
define $\delta^\ast(a)$ as the restriction of $\delta(a)$ on ${\cal 
H}^\ast$. Use a direct integral representation ${\cal H}_{\mu^\ast,N^\ast}$
of ${\cal H}^\ast$. Then, if $V^\ast$ denotes the corresponding unitary 
operator 
\be \label{0.62}
<\Psi,\delta(a) \Psi'>_{{\cal H}^\ast}:=\int \; d\mu^\ast(x)\;
\delta^\ast(x) <\psi(x),\psi'(x)>_{{\cal H}^\ast_x}
\ee
for all $\Psi,\Psi'\in \Phi$. It follows that if $\delta^\ast(x)$ is such 
that $\int d\mu^\ast(x) \delta^\ast(x) g(x)=g(0)$ for all measurable $g$ 
then 
\be \label{0.63a}
<\eta^\ast(\Psi),\eta^\ast(\Psi')>^\ast_{phys}=<\psi'(0),\psi(0)>_{{\cal 
H}^\ast_0}
\ee
so RAQ reproduces the results of the direct integral decomposition 
provided that $\{\psi(0);\;\Psi\in \Phi\}$ is dense in ${\cal 
H}^\ast_0$ and that there exist a rule for choosing representatives 
$x\mapsto \psi(x)$ for all $\Psi\in \Phi$ such that the numbers 
$<\psi(x),\psi'(x)>_{{\cal H}^\ast_x}$ are finite, at least at $x=0$. 
Notice that this 
issue about the representatives only arises in the construction of the 
physical inner product: For the elements $F\in\Phi^\ast$ 
the numbers $F[f]$ are of course well defined for any $f\in \Phi$,
however, the complication lies in the Rigging Map $\eta:\;\Phi\to 
\Phi^\ast_{phys}$ without which there is no physical inner product and 
which may 
actually not produce elements of $\Phi^\ast$ 
unless one has a rule for choosing appropriate representatives.
In other words, if one has an element $F\in \Phi^\ast$ which solves the 
constraints, then it may not be possible to display it in the form 
$F=\eta(f)$ for some $f\in \Phi$ unless one has resolved the issue about 
the representatives.\\
\\
Several remarks are in order:
\begin{itemize}
\item[i)] {\it Group Averaging}\\
Group averaging is a heuristic method to define the rigging map or,
in other words, the $\delta-$``operator `` $\delta(a)$.
For a single self -- adjoint constraint $C$ it 
consists of the formula
\be \label{0.63}
\eta(f):=\int_{-\infty}^\infty \; \frac{dt}{2\pi} \;<e^{it\hat{C}}f,.>
\ee
where the inner product indicated is the one on the unreduced Hilbert 
space. We now show that (\ref{0.63}) is wrong in general.
\begin{itemize}
\item[1.] {\it Purely Continuous Spectrum}\\
Consider the operator $C=p^2$. We have, using the momentum space 
representation
\be \label{0.64}
<\eta(f),\eta(f')>=\int_\Rl dp[\int_\Rl\frac{dt}{2\pi} e^{itp^2}]
\overline{f'(p)} f(p)=\lim_{p\to 0}\frac{\overline{f'(p)} 
f(p)}{|p|}
\ee
which is ill defined even for $f,f'$ in the dense subspace of functions of 
rapid decrease. The reason of failure is that in this case (\ref{0.63}) 
does not take into account the appropriate spectral measure $\mu$: 
We may choose $\Omega_1=\sqrt{\exp(-p^2/2)/\sqrt{2\pi}},\;
\Omega_2=p\Omega_1$. A straightforward calculation reveals that 
$\rho_1(x)=2/(1+x),\;\rho_2(x)=2x/(1+x)$ and 
$d\mu(x)=\frac{dx}{2\sqrt{2\pi x}}e^{-x/2}(1+x)$ where with 
$f=f_1(p^2)\Omega_1+f_2(p^2)\Omega_2$ we have 
\be \label{0.65}
<f,f'>=\int_{\Rl^+}\; d\mu(x)\;
[\rho_1(x)\overline{f_1(x)}f'_1(x)+
\rho_2(x)\overline{f_2(x)}f'_2(x)]
\ee
Since $\rho_1(0)=2>0,\;\rho_2(0)=0$ our prescription yields the correct 
result that the physical Hilbert space is one dimensional, isomorphic to 
$\Cl$, and can be thought of as the span of the vector $\Omega_1$.
\item[2.] {\it Purely Discrete Spectrum}\\
In the previous example one could rescue the proposal (\ref{0.63}) by 
selecting a ``reference vector'' $f_0$ and to formally define a new 
rigging map $\eta'(f):=\eta(f)/\eta(f_0)[f_0]$. This would take care of 
the singularity at $p=0$. We now show that even with this modification 
(\ref{0.63}) is completely wrong 
in the case of an entirely discrete spectrum. Take this time the 
harmonic oscillator $H=(p^2+q^2)/2$ with spectrum $x_n=\hbar(n+1/2),\;n\in 
\Nl_0$. Applying (\ref{0.63}) now yields on the eigenstates $e_n$ of the 
harmonic oscillator
\be \label{0.66}
<\eta(e_n),\eta(e_m)>_{Phys}:=\int_\Rl \frac{dt}{2\pi} 
<e_m,e^{it C} e_n>  
=\delta_{m,n} \delta(\hbar(m+1/2),0)=0
\ee
The physical Hilbert space would be empty because zero is not in the 
spectrum. Obviously we must normal order $C$ to remove the zero point 
energy, i.e. we quantize the quantum corrected classical expression 
$C':=(p^2+q^2)/2-\hbar/2$ which is semiclassically equivalent to $C$. But 
then 
the physical inner product diverges on the physical state $e_0$. Here 
one could repair the situation by integrating over the ``period''
$t\in [-\pi/\hbar,\pi/\hbar]$ but one sees already at this point that 
as compared to the case of the continuous spectrum the integration range 
cannot be chosen universally but depends on the spectrum of $C$. In 
particular, integrating over a finite period does not lead to the correct 
result in the case of a continuous spectrum.

However, one can think of even more generic situations. Consider the case 
of an operator with entirely discrete spectrum for which at least two 
eigenvalues are 
rationally independent. An example from LQG would be the area operator 
with a spectrum whose simplest eigenvalues are of the form 
$x=\ell_p^2\sum_p\sqrt{j_p(j_p+1)}$ where the sum runs over a finite set 
of points, the $j_p$ are half integral spin quantum numbers and $\ell_p^2$ 
is the Planck area.
Consider the family of operators $C_a=\mbox{Ar}(H)-a\ell_p^2$ where 
Ar$(H)$ is the 
area operator of an isolated horizon \cite{IH} and $a$ is a real number.
This kind of operators appear in the quantum entropy calculations in LQG.
The entropy is given by
\be \label{0.67}
S=\ln(\mbox{Tr}(P(a_0)),\;\;P(a_0):=
\sum_{a\in [a_0-1,a_0+1]} P_a
\ee
where $P_a$ is the projector onto the kernel of $C_a$. Now even if 
$a$ is in 
the spectrum of Ar$(H)$ it is {\it impossible} to define
$P_a$ via (\ref{0.63}) no matter how one chooses the period since 
$C_a$ has an infinite number of incommensurable eigenvalues. The 
way out here would be to
replace $\Rl$ by the Bohr compactification of the real line and $dt$ by 
the corresponding Haar measure. More in elementary terms one would define 
\be \label{0.68}
<f,P_a f'>:=\lim_{T\to \infty} \int_{-T}^T dt <f,e^{it C_a} f'>
\ee
However, this ergodic mean again does not lead to the correct result 
in the case of the continuous spectrum.
\item[3.] {\it Mixed spectrum}\\
Finally consider again the case of a mixed spctrum, e.g. the operator 
$C=C_1\otimes C_2$ on a Hilbert space ${\cal H}={\cal H}_1\otimes {\cal 
H}_2$ where $C_1$ has purely continuous spectrum such as in 1. and 
$C_2$ has purely discrete spectrum such as in 2. Now integrating over 
$\Rl$ in (\ref{0.63}) projects onto a physical Hilbert space which is 
isomorphic to the orthogonal complement of the ground state of 
$C_2$ in ${\cal H}_2$. Integrating over a finite period does not lead to 
any sensible result because the period of the eigenvalue $p^2 n\hbar$ 
of $C$ is $p$ dependent. Finally, ergodic averaging gives a physical 
Hilbert space isomorphic to ${\cal H}_1$. Hence in none of the cases 
does one recover the correct result which would be roughly isomorphic to 
the direct sum of ${\cal H}_1$ and ${\cal H}_2$ as we saw in section 
\ref{s0.5}.
\end{itemize}
We conclude that already in these simple examples group averaging only
leads to the correct physical result if one already knows the spectrum.
Even then it fails in the case of a mixed spectrum. Thus one must, similar 
as we observed in section \ref{s0.5}, first split the Hilbert space into 
its pure point and continuous part respectively. Again, this requires 
detailed knowledge of the spectrum so that rigorous methods more closely
tied to the spectral analysis of the operator such as the direct integral 
method suggest themselves.
\item[ii)] {\it Superselection Sectors, Non -- Amenable Groups and Group 
Averaging Constants}\\
In \cite{8.2} we find an example where group averaging has been carried 
out with respect to an infinite dimensional Lie group, the group of 
diffeomorphisms. With respect to the corresponding strong Dirac 
observables a certain 
superselection structure was discovered (these have a different origin 
than the separation between the types of spectrum discussed here). 
On each of those sectors 
the group averaging measures had to be carried out independently
for the following reasons: \\
1. The diffeomorphism group is not amenable, there is no
finite Haar measure on the diffeomorphism group. The only known Haar 
measures are counting measures.\\
2. In order to apply group averaging anyway one must renormalize the 
averaging procedure by formally dividing by the ``volume'' of the 
effective gauge group on each sector. 
The effective gauge group on a sector is the
subgroup of the diffeomorphism group each of whose elements has non -- 
trivial action on all vectors of the given sector.\\ 
This sector dependent volume is formally infinite and therefore the 
renormalized average is
only well defined up to a positive constant which could be different for 
each sector. This therefore leads
to a huge class of diffeomorphism invariant inner products depending on 
the choice of relative normalization constants between the group averaging 
measures of the respective sectors. It seems to be generic 
that group averaging leads to ambiguities, so -- called group averaging 
constants, for every system 
with superselection sectors and non -- amenable gauge groups. 
For the example 
in \cite{8.2} it is conceivable that these constants can be fixed by 
adding the weak Dirac observables to the analysis which is precisely what 
happened in the lower dimensional model of \cite{MMT}.

In contrast, the Master Constraint Programme in connection with DID,
where applicable, does
not lead to these ambiguities as we have seen, intuitively because the 
gauge group generated by the Master constraint is Abelean and hence 
amenable. We will see this explicitly in the examples of \cite{III}. In 
fact, the general analysis carried out in this paper did 
not depend at all on possible superselection sectors with respect to the
strong Dirac observables. 
\item[iii)] {\it Separability}\\
Notice that the Master Constraint 
Programme is not immediately applicable to the example of \cite{8.2} 
because the Hilbert space given there is not separable. However, that
Hilbert space is an uncountably infinite direct sum of separable Hilbert 
spaces which are invariant under the spatial diffeomorphism group and 
which are
labelled by diffeomorphism invariant continuous data (moduli) (that have 
to do with vertices of valence higher than four). Thus we may apply 
DID to each of these sectors separately, at least in principle, although 
this would be rather involved, see \cite{7.0,V}. Hence DID also 
applies 
to non -- separable Hilbert spaces which are (possibly uncountably 
infinite) direct sums of $\MCO-$invariant separable Hilbert spaces.
\item[iv)] {\it Continuity of the Dimension Function}\\
In \cite{Marolf}
the RAQ programme was also compared with the direct integral method
(called ``spectral analysis inner product'' there)
but the multiplicity function $N(x)$ was assumed 
to be constant in a neighbourhood of $x=0$. This is a somewhat 
reasonable assumption for the case of a single constraint. However, as we 
will see in the examples, not only 
do we not need this assumption but, moreover, the assumption is unphysical 
for the Master Constraint because in most examples the function $N$ is 
stronly discontinuous 
at $x=0$ as one should expect: This happens when, roughly speaking, 
the classical constraint manifolds $\MC-x=0$ have different dimension for 
different $x$. A typical example would be that $\MC=x>0$ is a (high 
dimensional) sphere but $\MC=0$ is a point. 
In quantum theory this is technically implemented by the fact that
while e.g. the functions $\rho^{ac}_n$ can be chosen to be continuous for 
all $n$, usually an infinite number of them are non-zero for $x\not=0$ but 
vanish at $x=0$. Thus, while the $\rho^{ac}_n(x)$ are actually continuous,
the function $N(x)$ is not.
\item[v)] {\it Direct Integrals and Rigged Hilbert Spaces}\\
In 
\cite{Marolf} it was already noticed that the choice of $\Phi$ is critical 
for the size of the resulting ${\cal H}^\ast_{phys}$ (the superselection 
structure corresponding to pure point and continuous spectrum was also 
noticed there). From that perspective it 
is surprising that the direct integral decomposition does not  
depend on the additional structure $\Phi$. In fact, modulo the 
prescription for how to choose the $\rho^c_n(x)$ for the continuous part 
of 
the spectrum, the physical Hilbert space does not need the structure of 
$\Phi$. Notice that the choice of the $\rho^\ast_n(x)$ is in one to one 
correspondence with the choice of representatives for the direct 
integral decomposition of a given
cyclic system of vectors $\Omega_n$, that is, with representatives 
$\omega_n(x)=\sqrt{\rho_n(x)} e_n$. 

However, it might be useful, at least for reasons of completeness, to 
relate the elements of $\cal H$ and of ${\cal H}_{Phys}$ and to display 
the relation between RAQ and the direct integral decomposition (DID).
In \cite{2.2} we 
find one method for how to do that, see also the summary in the appendix 
of \cite{7.0}. It uses the machinery of Rigged Hilbert Spaces and we 
recall here the essential steps of the construction, see 
\cite{2.2,7.0} for more details. We drop the label $\ast$ for simplicity,
focussing on one sector only for the remainder of this section.

A Rigged Hilbert 
Space consists of a so -- called Gel'fand triple 
$\Phi \hookrightarrow {\cal H} \hookrightarrow \Phi'$ consisting of a 
topological vector space $\Phi$ which is dense in $\cal H$ (in the 
topology of $\cal H$) and the topological (rather than algebraic) dual 
of $\Phi$ (continuous linear functionals). These spaces arise from a 
chain of separable Hilbert spaces $\Phi_N,\;N=1,2,..$ equipped with inner 
products
$<.,.>_N$ subject to the condition $||.||_N\le ||.||_{N+1}$ and  
$\Phi_{N+1}\subset \Phi_N$. One now defines $\Phi:=\cap_N \Phi_N$,  
equips it with the metric 
$d(F,F'):=\sum_N 2^{-N} ||F-F'||_N (1+||F-F'||_N)^{-1}$ and completes.
This makes $\Phi$ a Fr\'echet space, i.e. a complete, metrizable 
locally convex topological space\footnote{A topological space is called 
locally convex if its topology is defined by a family of seminorms 
separating the points. A seminorm is a norm just that the requirement 
$||F||=0 \;\Rightarrow \; F=0$ is dropped. A 
locally convex topological space is metrizable 
if and ond if its family of seminorms is countable.}.  
Such a structure is called a countably Hilbert space.  
One also defines $\Phi_{-N}:=\Phi'_N,\;n=1,2,..$ where $\Phi'_N$ is the 
topological 
dual of $\Phi_N$. (By the Riesz lemma, Hilbert spaces are reflexive and 
hence we may identify $\Phi'_N$ with $\Phi_N$.) 
From $\Phi_{N+1}\subset \Phi_N$ we conclude that any $l\in \Phi'_N$ is 
also an element of $\Phi'_{N+1}$, hence $\Phi'_{-N}\subset\Phi'_{-(N+1)}$.
One now defines $\Phi':=\cup_N \Phi'_N$. 
A nuclear space is a countably Hilbert space such that for all $M$ there 
exists $N\ge M$ and such that the natural embedding 
$T_{NM}:\Phi_N\to \Phi_M$ is trace class (nuclear) , i.e. if $B^{(N)}_k$ 
is an orthonormal basis of $\Phi_N$ then 
$T_{NM} F=\sum_k \lambda_k <B^{(N)}_k,F>_N B^{(M)}_k$ where 
$\sum_k \lambda_k<\infty,\;\lambda_k\ge 0$. 
Finally 
one needs an inner product $<.,.>_0$ and defines $\Phi_0:={\cal H}$ as the 
completion of $\Phi$ in that inner product. One can show that one obtains 
a chain of Hilbert 
spaces $\Phi_{N+1}\subset \Phi_N,\;N\in \Zl$. 
A Rigged Hilbert space is such a chain of Hilbert spaces such that 
convergence in the topology of 
$\Phi$ implies convergence in the topology of $\cal H$. \\
\\
To see that this structure 
is naturally available in the context of the Master Constraint Programme, 
notice that every 
self-adjoint operator $\MCO$ on a Hilbert space has a dense set $\cal D$ 
of 
$C^\infty-$vectors 
of the form $\Omega_g=\int_\Rl\; dt\; g(t)\exp(it\MCO)\Omega$ where 
$g\in C^\infty_0(\Rl)$ is a smooth function of compact support and 
$\Omega\in {\cal H}$ and one computes $\MCO^n \Omega_g=(-1)^n 
\Omega_{g^{(n)}}$. 
We now define for $F,F'\in {\cal D}$ 
\be \label{0.69}
<F,F'>_N:=\sum_{k=0}^N <F,\MCO^k F'>_{{\cal H}}
\ee
which defines positive semi definite inner products {\it because $\MCO$ 
is positive semi definite}. We define $\Phi_N,\;N=0,1,2..$ as the Cauchy 
completion of ${\cal D}$ in the norm defined by (\ref{0.69}) and see that 
the conditions on a countably 
Hilbert space are satisfied. Moreover, using a direct integral 
decomposition of $\cal H$ and a cyclic system of vectors $\Omega_n$ 
we find that 
\be \label{0.70}
||F||_N^2=\int_{\Rl_+}\; d\mu(x)\; \frac{x^N-1}{x-1} \;
\sum_n\rho_n(x) |F_n(x)|^2
\ee
where $F=\sum_n F_n(\MCO) \Omega_n$
which shows that the norms of $F$ grow rapidly with $N$ provided that 
$\MCO$ is an unbounded operator. This typiically implies that $\Phi$ will 
be a nuclear
space because, suppose that $B^{(0)}_k$ is an orthonormal basis of 
${\cal H}$ constructed from functions in $\cal D$ by the Gram -- Schmidt 
algorithm. This means we may choose 
\be \label{0.71}
\sum_n \rho_n(x) \overline{b^{(0)}_{kn}(x)} b^{(0)}_{k'n}(x)=
\delta_{kk'} \sigma_k(x)
\ee
$\mu-$a.e for some measurable function $\sigma_k$ where 
$B^{(0)}_k=\sum_n b^{(0)}_{kn}(E)\Omega_n$. This means that 
$B^{(N)}_k:=\sqrt{\mu_{Nk}}^{-1} B^{(0)}_k$ is 
an 
orthonormal basis of $\Phi_N$ where 
\be \label{0.72}
\mu_{kN}:=\int d\mu(x) \sigma_k(x)\frac{x^N-1}{x-1} 
\ee
If these moments of $\mu$ grow sufficiently fast then by the completeness
relation for $F\in \Phi_N$
\be \label{0.73}
T_{NM} F=\sum_k <B^{(N)}_k,F>_N B^{(N)}_k
=\sum_k \sqrt{\frac{\mu_{kM}}{\mu_{kN}}} <B^{(N)}_k,F>_N B^{(M)}_k 
\ee
and we identify $\lambda_k=\sqrt{\frac{\mu_{kM}}{\mu_{kN}}}$ which will 
satisfy the trace class condition depending on the growth of the 
moments. Notice that also the compatibility condition between the 
topologies of $\Phi$ and $\cal H$ is automatically stisfied here 
because convergence in the topology of $\Phi$ implies convergence with 
respect to all the $||.||_N$ in particular $||.||_0=||.||_{{\cal H}}$.\\
\\
In any case, if a Rigged Hilbert Space Structure is available, one has the 
following result:
\begin{Theorem} \label{th0.10} ~~\\
Let $\Phi\hookrightarrow {\cal H} \hookrightarrow \Phi'$ be a 
separable Rigged 
Hilbert Space. Let ${\cal H}\equiv {\cal H}_{\mu,N}$ be a direct integral 
representation of ${\cal H}$ subordinate to a self -- adjoint operator
$\MCO$.\\
i)\\
Then there exists a nuclear operator 
$T_x:\;\Phi\to {\cal H}^\oplus_x$ such that $T_x F=f(x)$ $\mu-$a.e.
where $(f(x))_{x\in \Rl^+}$ is the direct integral representation of 
$F\in \Phi$. Moreover, the norms $||T_x f||_{{\cal H}_x}$ of the  
the vector valued function 
$x\mapsto T_x F$ are uniquely 
defined through the operator $T_x$ (and not only $\mu-$a.e.). \\
ii)\\
The maps
\be \label{0.74}
\eta_x:\;\Phi\to \Phi';\;F\mapsto \eta_x(F),\;\;
\eta_x(F)[F']:=<T_x F,T_x F'>_{{\cal H}^\oplus_x}
\ee
are Rigging Maps and $\eta_0$ defines the physical Hilbert space.
Moreover, the subset of $\Phi'$ defined by the images under the $\eta_x$ 
defines a complete set of generalized eigenvectors, that is, 
$\eta_x(F)[\MCO F']=x\eta_x(F)[F']$ for all $F,F'\in \Phi$ and 
$\cup_x \eta_x(\Phi)\subset \Phi'$ separates the points of $\Phi$.
\end{Theorem}
Actually in order to prove the theorem it is sufficient that $\Phi$ is a 
Fr\'echet space and that there exists a map $T=T_2 T_1$ where 
$T_1:\;\Phi\to {\cal H}_1$ is a continuous embedding of $\Phi$ into some 
Hilbert space ${\cal H}_1$ and $T_2$ is a nuclear operator. To prove the 
theorem one proceeds as follows: Given a direct integral representation of 
${\cal H}$ and an orthonormal basis $B_k$ of ${\cal H}$ with direct 
integral representation $(b_k(x))_{x\in \Rl}$ one {\it chooses} for each
$x$ a representative $b_k(x)$ once and for all. This means that on top 
of choosing values for the $\rho_n(x)$ which define the ${\cal 
H}^\oplus_x$ we must also choose values for the measurable functions 
$b_{kn}(x)$ where $B_k=\sum_n b_{kn}(E)\Omega_n$. 
The nuclear structure enables one to show 
that there exists an $N$, independent of $x$, $\lambda_k\ge 0$  
with $\sum_k\lambda_k<\infty$ and an 
orthonormal basis 
$B^{(N)}_k$ of $\Phi_N$ such that 
the operator 
\be \label{0.75}
T_x F:=\sum_k \lambda_k <B^{(N)}_k,F>_N b_k(x)=
\sum_n \sqrt{\rho_n(x)}[\sum_k \lambda_k <B^{(N)},F>_N b_{kn}(x)] e_n
\ee
is nuclear (where $e_n,\;\rho_n(x)>0$ is an orthonormal basis of 
${\cal H}^\oplus_x$) and 
coincides with $f(x)$ $\mu-$a.e. More precisely, the norm 
of (\ref{0.75}) converges $\mu-$a.e. to a finite number and $T_x F$
is then defined {\it everywhere} by setting it to zero at those $x$ for 
which the norm of 
(\ref{0.75}) does not converge.\\
\\
What has been gained?\\
Notice that the theorem supposes that one has already chosen the Hilbert
spaces ${\cal H}^\oplus_x$, the ambiguity in the $\rho_n(x)$ has been 
fixed by making a definite choice. Furthermore, one must make a choice of 
the bases $B_k, B^{(N)}_k$ and one must choose representatives 
$b_k(x)$. One will make the cyclic system consisting of the $\Omega_n$ 
part of the basis $B_k$ so that $b_{kn}(x)=\delta_{mn}$ for $B_k=\Omega_n$ 
but clearly there are more $B_k$ than $\Omega_n$. 
This is more than one has to choose in order to define the ${\cal 
H}^\oplus_x$. The only advantage of this theorem is that once one has made 
these choices to define the $T_x$, one can assign definite numbers to 
the inner products $<T_x F,T_x F'>_{{\cal H}_x}$ or in other words one can
reduce the ambiguity in choosing the representative $f(x)$ universally
(i.e. independently of $F\in \Phi$). 

Hence we see that RAQ, even in its precise form given in theorem 
\ref{th0.10}, is more ambiguous than DID. This seems surprising in view of 
the fact that using the Rigged Hilbert space based on the nuclear space of 
test functions of rapid decrease on $\Rl$ and a direct integral 
representation of 
$L_2(\Rl,dx)$ subordinate to the momentum operator the 
$T_x F$ are just the Fourier coefficients 
of the Fourier integral defining $F$ in the sense of $L_2$ functions 
and that these coefficients are naturally smooth and of rapid decrease 
again. However, notice that secretly one has made also choices here, 
nobody can prevent one to make those Fourier coefficients arbitrarily 
discontinuous on a countable subset of $\Rl$.
Moreover, in the case of the Fourier transform one has actually a cyclic
vector for the momentum operator(s) and therefore the $\rho_n$ are 
naturally identical to the constant function equal to one. In general 
little seems to be known in the mathematical literature about the 
connection between the ``natural'' continuity of the $T_x F$ and the 
choice of $\Phi$.  
\end{itemize}
In summary we see that DID not only extends RAQ
to the case of structure functions, it can also be used in order to 
reduce the ambiguity even for the rigorous version of RAQ since one is not 
in the need to choose the additional structures 
$\Phi, T_x$. These additional structures are of no interest whatsoever 
to the direct integral decomposition because ${\cal H}^\oplus_0$ is 
determined by 
independent means, subject to the physical prescription of enforcing 
an irrducible, self -- adjoint representation of a complete subalgebra of 
all Dirac 
observables with a good semiclassical limit which fixes the values of the 
$\rho_n(0)$ as much as it 
possibly can\footnote{The additional structure $\Phi,T_x$ might be 
helpful,
however, in order to fix the values of the measurable functions 
$G_{mn}(x)$ for all $x$, not only for $x=0$, although this is not 
particularly interesting from the point of view of the physical Hilbert 
space.}.  
The physical Hilbert space is then isomorphic to the $\ell_2$ 
space 
of sequences $(z_n)$ for which $\sum_n \rho_n(0) |z_n|^2$ converges. 
We do not need to worry about the question whether the $z_n$ can be 
thought of as the values $\Psi_n(0)$ of the measurable functions 
$\Psi_n$ defined by $\Psi=\sum_n \Psi_n(a)\Omega_n$ and whether there
is a natural representative $x\mapsto \Psi_n(x)$ in the corresponding 
equivalence class. Finally, 
there is no need to deal with the formal ``operator'' $\delta(a)$.

\section{Algorithmic Description of the Direct Integral Decomposition}
\label{s9}

Given a self -- adjoint operator $\MCO$ on a separable Hilbert space
$\cal H$, the Direct Integral Decomposition (DID) method to solve the 
constraint
$\MCO=0$ consists of the following steps:
\begin{itemize}
\item[Step I.] {\it Spectral Measures}\\
For any Lebesgue measurable set $B$ and any $\Psi\in {\cal H}$ determine 
the spectral measures 
$\mu_\Psi(B):=<\Psi,E(B)\Psi>$ where $E$ is the p.v.m. underlying 
the 
operator $\MCO$, that is, $E(B)=\chi_B(\MCO)$ where $\chi_B$ is the 
characteristic function of the set $B$. It is actually sufficient to 
construct these for the sets $B_x:=(-\infty,x]$ and we set 
$\mu_n(x):=\mu_n(B_x)$. To construct the $\mu_\Psi(x)$ explicitly
from a given operator $\MCO$ is very hard in general, however, if 
one can construct the bounded resolvent  
$R(z):=(\MCO-z{\bf 1})^{-1}$ 
for $\Im(z)\not=0$ which one can often determine by Green function 
techniques then one can use Stone's formula
\be \label{0.76}
\frac{1}{2}(E([a,b])+E((a,b))=s-\lim_{\epsilon \to 0}
\int_a^b dt [R(t+i\epsilon)-R(t-i\epsilon)]
\ee
In fortunate cases the operator $\MCO$ is of the form 
$F(\{\hat{a}_\alpha\})$
where the $\hat{a}_\alpha$ form a mutually
commuting set of other self -- adjoint operators for which 
one knows explicitly a representation as multiplication operators on a 
space of square integrable functions in some variables $y_\alpha$. Then 
${\cal H}$ can be represented as some $L_2$ space on the space of 
$y_\alpha$ and $\MCO$ acts by multiplication by $F(\{y_\alpha\})$. 
The same applies when some of the $\hat{a}_\alpha$ have discrete spectrum 
in 
which case the $L_2$ space is replaced by an $l_2$ space and $y_\alpha$
by the corresponding eigenvalue.
\item[Step II.] {\it Separation of Discrete and Continuous Spectrum}\\
We say that $\mu_\Psi$ is of pure point or continuous type respectively
if $\mu_\Psi$ has support on a discrete set of points or is not supported 
on one point sets respectively. Let ${\cal H}^{pp},\;{\cal H}^c$ be the 
completion of the linear span of vectors such that $\mu_\Psi$ is of the 
respective type. Then it is always true that ${\cal H}={\cal H}^{pp}\oplus 
{\cal H}^c$.
\item[Step III.] {\it Cyclic System}\\
For each sector ${\cal H}^\ast,\;\ast=\{pp,c\}$ determine a minimal sysyem 
of mutually orthogonal $C^\infty-$vectors, that 
is, normalized vectors $\Omega^\ast_n,\; n=1,..,M^\ast\le \infty$ such 
that\\
A. all powers of $\MCO$ are defined on $\Omega^\ast_n$,\\  
B. $<\Omega^\ast_m,\MCO^N\Omega^\ast_n>=0$ for all $N=0,1,2,..$ and all 
$m\not=n$,\\
C. the finite linear span of the $\MCO^N \Omega^\ast_n$ is dense in 
${\cal H}^\ast$,\\
D. $M^\ast$ cannot be reduced without violating condition C.\\
In practice one starts from a known orthonormal basis $B_k$ for $\cal H$
and tries to identify elements $\Omega^\ast_n$ of that basis such that any 
$B_k$ is a finite 
linear combination of the $\MCO^N \Omega^\ast_n$. Different choices of 
$\Omega^\ast_n$ satisfying A. -- D. lead to unitarily equivalent direct 
integral decompositions ${\cal H}^\ast_{\mu^\ast,N^\ast}=\int \; 
d\mu^\ast(x) 
\;{\cal H}^{\ast \oplus}_x$ of ${\cal H}^\ast$ and in fact the measure 
class of $\mu^\ast$
and the dimension $N^\ast(x)$ of ${\cal H}^{\ast\oplus}_x$ is unique 
$\mu^\ast-$a.e.
(independent of the $\Omega^\ast_n$). The minimal set of $\Omega^\ast_n$ 
is such 
that the supports $S^\ast_n$ of the functions $\rho^\ast_n$ defined below 
are ordered such that $S^\ast_{n+1}\subset S^\ast_n$.  
%
\item[Step IV.] {\it Total Measure and Radon Nikodym Derivatives}\\
Let $\mu^\ast_n(B):=<\Omega^\ast_n,E(B)\Omega^\ast_n>$ and 
define $\mu^\ast(x):=\sum_n c_n^\ast \mu^\ast_n(x)$
where where $c^\ast_n>0,\; \sum_n c^\ast_n=1$. Its
measure class is actually unique (indendent of the choice of 
$\Omega^\ast_n,\;c^\ast_n$). Let, if the limit exists  
\be \label{0.77}
\rho^\ast_n(x):=\lim_{y\to 0+} 
\frac{\mu^\ast_n(x+y)-\mu^\ast_n(x-y)}{\mu^\ast(x+y)-\mu^\ast(x-y)} 
\ee
For $\ast=pp$ the numbers (\ref{0.77}) always exist, they are always non 
-- negative and the number $N^{pp}(x)$ of those $n$ for which 
$\rho^{pp}_n(x)$ does not vanish is 
independent of the $\Omega^{pp}_n$. 
For $\ast=c$ they are granted to exist only $\mu^c-$a.e., are naturally
non -- negative (but might be infinite) and the number $N^c(x)$ is only 
unique $\mu^c-$a.e. (i.e. independent of the $\Omega^\ast_n, c^\ast_n$).   
For a given system $\Omega_n^\ast,\;c^\ast_n$ fix these constants once and 
for all by choosing a representative subject to the requirement 
that the resulting physical Hilbert space is an irreducible self -- 
adjoint representation of a complete subalgebra of all Dirac observables 
and admits 
a sufficient number of semiclassical states. For any other choice of 
$\Omega_n^\ast,c_n^\ast$ these constants are then fixed by the requirement 
that the induced representation of the strong Dirac observables is 
unitarily equivalent to the given one, see below.
In practice, it is often sufficient to  
choose a representative, if it exists, which is continuous from the right 
at $x=0$ and to set $\rho^c_n(0):=\lim_{x\to 0+} \rho^c_n(x)$. If it does 
not exist, set $\rho^c_n(0)=0$ if also not fixed by the aforementioned 
physical criteria. 
\item[Step V.] {\it Physical Hilbert Space}\\
Let $e^\ast_n,\;n=1,2,..,M^\ast$ be an orthonormal basis in some abstract 
Hilbert space. The Hilbert space ${\cal H}^\oplus_x$ is the 
space of vectors of the form $\sum_{\ast=\{pp,c\}} 
K_\ast
\sum_{n=1}^{M^\ast} 
\sqrt{\rho^\ast_n(x)} z^\ast_n e^\ast_n$ with $z_n^\ast\in \Cl$ for which 
the norm squared 
$K_{pp} \sum_n \rho^{pp}(x) |z^{pp}_n|^2+K_c \sum_n \rho^c(x) |z^c_n|^2$
converges. 
The positive constants $K_\ast$ can possibly be determined if there are 
weak Dirac observables which mix the continuous and discrete part of the 
Hilbert space in which case they must be chosen so that the algebra of 
weak Dirac observables are also represented self -- adjointly.
The physical Hilbert space coincides with ${\cal H}^\oplus_{x=0}$. 
\item[Step VI.] {\it Representation of Dirac Observables}\\
We will restrict the discussion to strong Dirac observables. See
section \ref{s0.6} for the more interesting case of weak Dirac 
observables.
A strong bounded Dirac observable $D$ commutes with $\MCO$ and then 
preserves the 
fibres ${\cal H}^{\ast\oplus}_x$. In terms of a cyclic $C^\infty$ 
system $\Omega^\ast_n$ we 
find measurable functions $G^\ast_{mn}$ such that 
$D\Omega^\ast_m=\sum_n G^\ast_{mn}(\MCO) \Omega^{\ast}_n$.
Then 
\be \label{0.79a}
D(x) e^\ast_m=\sum_n D^\ast_{mn}(x) e^\ast_n,\;\;
D^\ast_{mn}(x)=
\chi_{M^\ast(x)}(m)\chi_{M^\ast(x)}(n) 
\sqrt{\frac{\rho^\ast_n(x)}{\rho^\ast_m(x)}} G^\ast_{mn}(x)
\ee
where $M^\ast(x)=\{n\in \Nl;\;\rho^\ast_n(x)>0\}$ and a representative 
for $G^\ast_{mn}(x)$ was chosen. For symmetric choice of 
$\rho^\ast_n(x) G^\ast_{mn}(x)$ expression 
(\ref{0.79a}) is automatically self -- adjoint and bounded on ${\cal 
H}_{phys}$ if 
$D$ is on ${\cal H}$ no matter how the $\rho^\ast_n(0)$ are chosen. 
Moreover, for different choices of $\Omega^\ast_n, 
c^\ast_n$ we can always choose the 
corresponding different $\rho^\ast_n(0), G^\ast_{mn}(0)$ such 
that the induced 
representations of the strong Dirac observables on the physical 
Hilbert space are unitarily equivalent, see above. This makes in 
particular the 
dimension $N^\ast(0)$ of ${\cal H}^{\ast\oplus}_0$ independent of
the $\Omega^\ast_n,\;c^\ast_n$.  
Thus we need to choose the $\rho^\ast_n(0), G^\ast_{mn}(0)$ only once
and then
only the weak Dirac observables can fix the ambiguity in the 
$K_\ast$. The $\rho^\ast_n(0)$ should be constrained by 
the requirement that the physical Hilbert space contains a sufficient 
number of semiclassical states and the requirement that
the physical Hilbert space should be an irreducible representation for the 
induced action of a complete subalgebra of bounded self -- adjoint Dirac 
observables which are defined 
on the unconstraned Hilbert space $\cal H$, see above.
\end{itemize}

\section{Conclusions}
\label{s10}

In our companion papers \cite{II,III,IV,V} we will apply the 
Master Constraint Programme and in particular the Direct Integral 
Decomposition (DID) to 
models of varying degree of complexity, starting with finite dimensional 
systems with a finite number of Abelean first class constraints linear in 
the momenta and ending with infinite dimensional systems (interacting 
field theories) with an infinite number of first class constraints not 
even polynomial in the momenta which close with structure functions only 
rather than with structure constants. This latter worst case scenario is 
precisely the case of 3+1 dimensional General relativity plus matter and 
therefore we believe that the Master Constraint Programme has been
scrutinized in sufficiently complicated situations.

We hope to be able to convince the reader that the $\MCPW$ can be very
successfully applied to this wide range of constrained theories including
those where other methods fail. While due care must be taken when squaring
constraints, the Master Constraint Programme is sufficiently flexible in 
order to deal with
the associated factor ordering problems and the worsened ultraviolet
behaviour in these examples.

The $\MCPW$ was invented in \cite{8.1} in order to overcome the
present obstacles in implementing the dynamics in LQG outlined
in the introduction and in more detail in \cite{7.0}. That it also
provides the physical inner product of theory and even offers some handle 
on the Dirac observables comes at quite a surprise. So far the only 
systematic procedure in order
to arrive at the physical inner product of a constrained theory was the
group averaging method of RAQ \cite{1.10} reviewed in section 
\ref{s0.7} and RAQ ideas were used very
successfully in LQG in order to derive the Hilbert space of spatially
diffeomorphism invariant states \cite{8.2}. However, the RAQ method
usually fails when the constraint algebra is not a finite dimensional
Lie algebra.

It is here where the Master Constraint Programme can take over and enables 
us
to make progress, since the (infinite) dimensional constraint algebra,
whether it closes with structure functions or structure constants, is
replaced by a one -- dimensional Abelean Lie algebra. The only restriction 
is that the Hilbert space be separable, or that it can be decomposed into
a possibly uncountably infinite direct sum of $\MCO-$invariant separable
Hilber spaces. For this case
the RAQ method and the $\MCPW$ essentially coincide with the difference
that the $\MCPW$ does not require the additional input of a nuclear
topology on a dense subspace of the kinematical Hilbert space, the
Master Constraint Programme just uses the spectral theory of the $\MCO$ 
and other
physically motivated structures outlined in section \ref{s0.5}
which involve less ambiguities than in the RAQ programme as shown in 
section \ref{s0.7}. On the other hand
the $\MCOW$ itself generically provides a natural nuclear topology
\cite{2.2} as explained in \cite{7.0} and also in section \ref{s0.7} so 
that in this case both procedures are really rather close.

The success of the programme of course rests on the question whether
we can really quantize 3+1 General Relativity (plus matter) with this
method. While we will show in \cite{8.1} that there are no more
mathematical obstacles on the way in order to complete the programme,
it is not yet clear whether the resulting physical Hilbert space contains
a sector which captures the semiclassical regime of both General
Relativity and Quantum Field Theory on curved spacetimes. In order to show
this, it will be necessary to develop approximation methods to construct
the physical Hilbert space, such as path integral methods using coherent
states, thus defining a new type of spin foam models as outlined in
\cite{7.0}. Approximation methods are mandatory because General Relativity 
is a fantastically difficult
interacting quantum field theory with no hope to be solvable exactly. In
order
to complete this step we must develop spatially diffeomorphism invariant
coherent states because the Master constraint is defined only on the 
spatially diffeomorphism invariant Hilbert space. 
After having constructed the physical Hilbert space by DID methods,
at least approximately,
one must eventually construct physical semiclassical states. All of these
steps are parts of a hard but, we believe, not hopeless research project
which is now in progress. At least the mathematical obstacles concerning
the solution of the Hamiltonian constraints are now out of the way and
we can in principle carry out the mathematical quantization programme to
the very end.\\
\\
The results of this series of papers, in our mind, demonstrate that the 
mathematics of the Master Constraint Programme succeeds in a large class 
of typical examples
to capture the correct physics so that one can be hopeful to be able to
do the same in full 3+1 quantum gravity.\\
\\
\\
\\
{\large Acknowledgements}\\
\\
We would like to thank Abhay Ashtekar and, especially, Hanno Sahlmann
for fruitful discussions.
BD thanks the German National Merit Foundation for financial support.
This research project was supported in part by a grant from 
NSERC of Canada to the Perimeter Institute for Theoretical Physics.

\end{document}